%% file: main-automatica.tex
\documentclass[twocolumn]{autart}

\usepackage[hidelinks]{hyperref}
\edef\endfrontmatter{\unexpanded\expandafter{\endfrontmatter}\noexpand\endNoHyper}

\usepackage{booktabs}
\usepackage{amsfonts}
\usepackage{nicefrac}
\usepackage{xcolor}

\usepackage{graphicx}
\usepackage[compress]{natbib}
\usepackage{mathtools,amsmath,amssymb}
\usepackage{enumitem}
\usepackage{etoolbox}
\AtEndEnvironment{pf}{\null\hfill$\square$}%
\usepackage{wrapfig}

\usepackage{algorithm}
\usepackage{algorithmic}

\theoremstyle{plain}
\newtheorem{theorem}{Theorem}[section]
\newtheorem{proposition}[theorem]{Proposition}
\newtheorem{lemma}[theorem]{Lemma}
\newtheorem{corollary}[theorem]{Corollary}
\theoremstyle{definition}

\theoremstyle{remark}
\newtheorem{remark}[theorem]{Remark}

\usepackage{abbreviations}
\usepackage{aligned-overset}
\usepackage{siunitx}
\usepackage{multirow}
\usepackage[labelformat=simple]{subfig}

\usepackage{tikz,pgfplots}
\pgfplotsset{compat=newest}

\usetikzlibrary{calc,fit,math,positioning}
\pgfplotsset{select coords between index/.style 2 args={
    x filter/.code={
        \ifnum\coordindex<#1\fi
        \ifnum\coordindex>#2\fi
    }
}}

\allowdisplaybreaks

\renewenvironment{thebibliography}[1]{%
    \begin{oldthebibliography}{#1}%
    \setlength{\parskip}{0ex}%
    \setlength{\itemsep}{0ex}%
}%
{%
    \end{oldthebibliography}%
}
\begin{document}
\setlength{\abovedisplayshortskip}{0.6ex plus1ex minus1ex}
\setlength{\abovedisplayskip}{0.6ex plus1ex minus1ex}
\setlength{\belowdisplayshortskip}{0.9ex plus1ex minus1ex}
\setlength{\belowdisplayskip}{0.9ex plus1ex minus1ex}
\begin{frontmatter}

\title{SafEDMD: A Koopman-based data-driven controller design framework for nonlinear dynamical systems\thanksref{footnoteinfo}}

\thanks[footnoteinfo]{
   F.\ Allgöwer is thankful that this work was funded by the Deutsche Forschungsgemeinschaft (DFG, German Research Foundation) under Germany's Excellence Strategy -- EXC 2075 -- 390740016 and within grant AL 316/15-1 -- 468094890.
    K.\ Worthmann gratefully acknowledges funding by the German Research Foundation (DFG; project number 507037103).
    R.\ Strässer thanks the Graduate Academy of the SC SimTech for its support.
    \\
    \phantom{11}\emph{Email address:}\\
    \texttt{straesser@ist.uni-stuttgart.de} (R. Strässer),\\
    \texttt{manuel.schaller@math.tu-chemnitz.de} (M. Schaller),\\
    \texttt{karl.worthmann@tu-ilmenau.de} (K. Worthmann),\\
    \texttt{berberich@ist.uni-stuttgart.de} (J. Berberich),\\ 
    \texttt{allgower@ist.uni-stuttgart.de} (F. Allgöwer).
}

\author[1]{Robin Str\"asser},
\author[2,3]{Manuel Schaller},
\author[2]{Karl Worthmann},
\author[1]{Julian Berberich},
\author[1]{Frank Allg\"ower}

\address[1]{
    University of Stuttgart, Institute for Systems Theory and Automatic Control, 70550 Stuttgart, Germany
}
\address[2]{
    Technische Universit\"at Ilmenau, Institute of Mathematics, 98693 Ilmenau, Germany
}
\address[3]{
    Chemnitz University of Technology, Faculty of Mathematics, 09111 Chemnitz, Germany
}

\begin{abstract}                
    The Koopman operator serves as the theoretical backbone for machine learning of dynamical control systems, where the operator is heuristically approximated by extended dynamic mode decomposition (EDMD). 
    In this paper, we propose SafEDMD, a novel stability- and feedback-oriented EDMD-based controller design framework. Our approach leverages a reliable surrogate model generated in a data-driven fashion in order to provide closed-loop guarantees. 
    In particular, we establish a controller design based on semi-definite programming with guaranteed stabilization of the underlying nonlinear system. 
    As central ingredient, we derive proportional error bounds that vanish at the origin and are tailored to control tasks.
    We illustrate the developed method by means of several benchmark examples and highlight the advantages over state-of-the-art methods.
\end{abstract}


\end{frontmatter}

\input{sec1-introduction}
\input{sec2-Koopman-operator}
\input{sec3-learning-architecture}
\input{sec4-controller-design}
\input{sec5-numerical-examples}
\input{sec6-conclusion}

\appendix

\input{secAppendix-proof-proportional-bound-DD-surrogate}
\input{secAppendix-proof-controller-design}
\input{secAppendix-proof-controller-continuous-time}

\bibliography{literature}

\clearpage
\bgroup
\setlength{\columnsep}{4pt}
\small
\begin{wrapfigure}[10]{l}{0.9in}
	\includegraphics[width=0.9in,height=1.15in,clip,keepaspectratio]{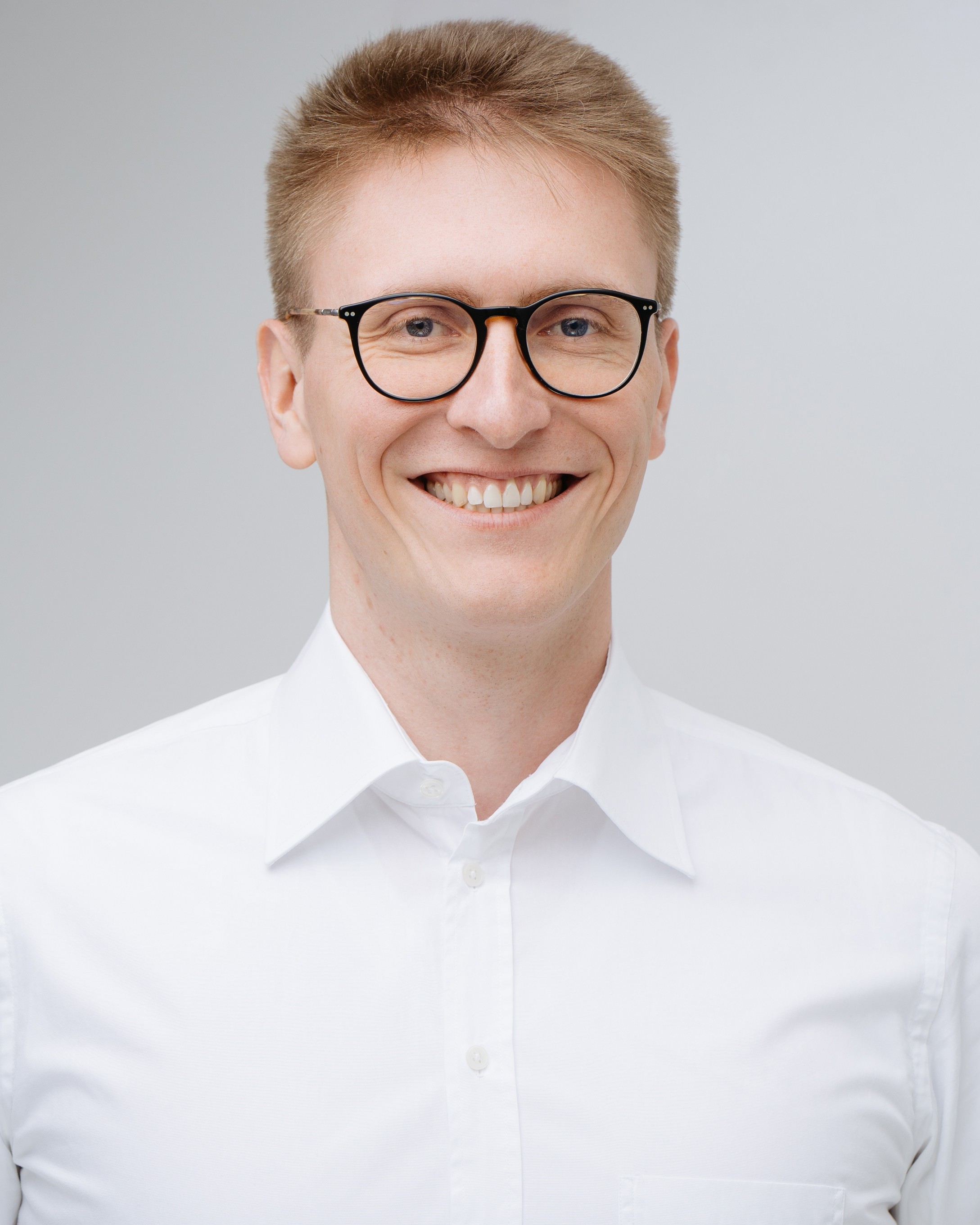}
	\vspace{-12pt}
\end{wrapfigure}
\noindent
\textbf{Robin Str\"asser} received a master’s degree in Simulation Technology from the University of Stuttgart, Germany, in 2020. Since 2020, he has been a Research and Teaching Assistant with the Institute for Systems Theory and Automatic Control and a member of the Graduate School Simulation Technology at the University of Stuttgart. His research interests include data-driven system analysis and control, with a focus on nonlinear systems. Robin Strässer received the Best Poster Award at the International Conference on Data-Integrated Simulation Science (SimTech2023) and the Best Paper Award at the European Robotics Forum (ERF2025).

\begin{wrapfigure}[10]{l}{0.9in}
	\vspace{-8pt}
	\includegraphics[width=0.9in,height=1.15in,clip,keepaspectratio]{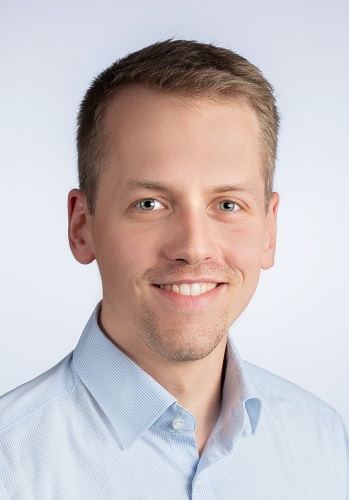}
	\vspace{-12pt}
\end{wrapfigure}
\noindent
\textbf{Manuel Schaller} obtained the M.Sc.\ and Ph.D.\ in Applied Mathematics from the University of Bayreuth in 2017 and 2021 respectively. From 2020-2023 he held a Lecturer position and a Junior Professorship at Technische Universität Ilmenau, Germany. Since August 2024, he is tenure track assistant professor at Chemnitz University of Technology, Germany. His research focuses on data-driven control with guarantees, port-Hamiltonian systems and efficient numerical methods for optimal control.
For his research he has been named junior fellow of the GAMM (Society for Applied Mathematics and Mechanics), received the Best Poster Award at the workshop on systems theory and PDEs (WOSTAP 2022) and is an elected member of the Young Academy of the European Mathematical Society (EMS) in 2024-2027.

\begin{wrapfigure}[10]{l}{0.9in}
	\vspace{-8pt}
	\includegraphics[width=0.9in,height=1.15in,clip,keepaspectratio]{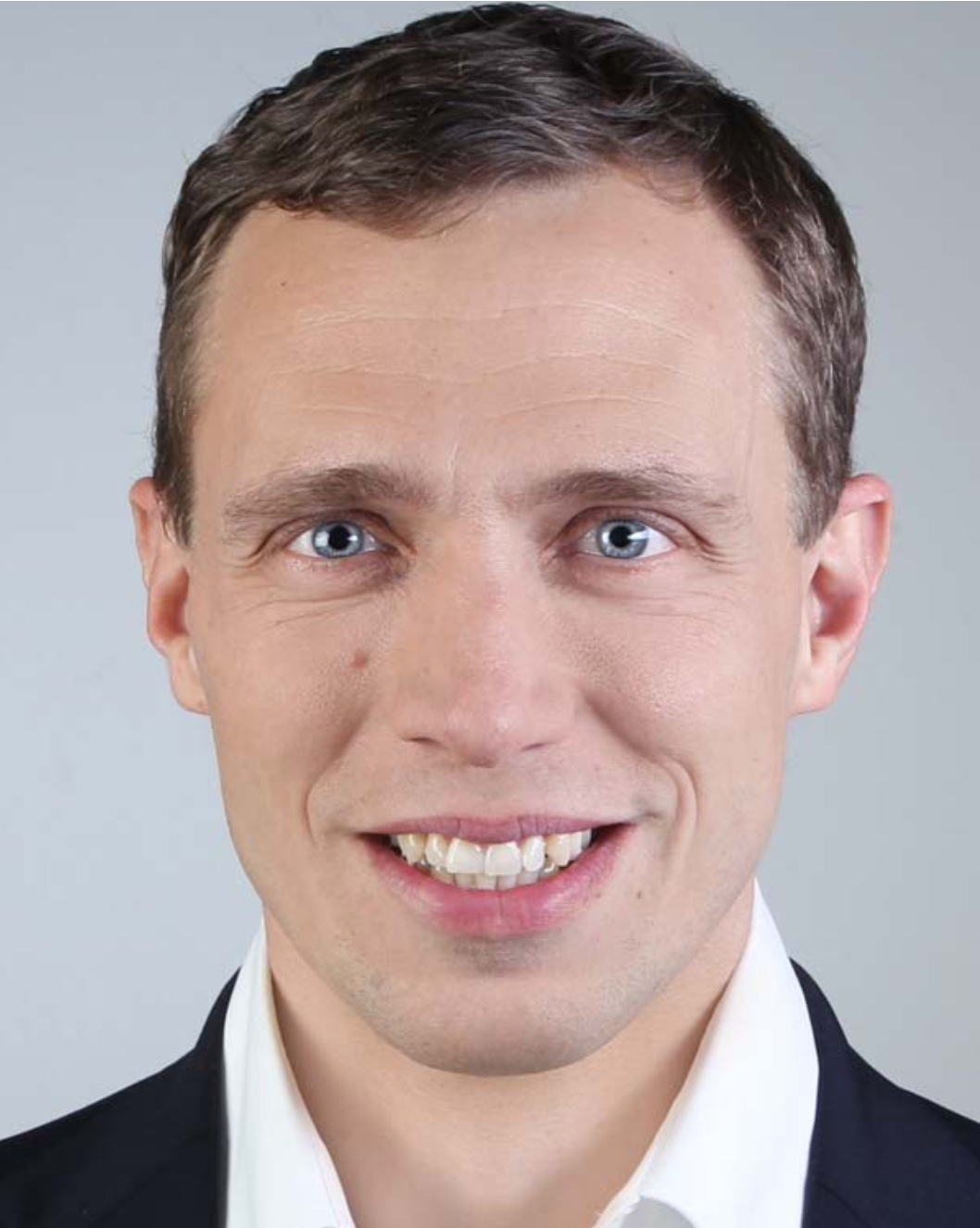}
	\vspace{-12pt}
\end{wrapfigure}
\noindent
\textbf{Karl Worthmann} received his Ph.D.\ in mathematics from the University of Bayreuth, Germany, in 2012. 2014 he become assistant professor at Technische Universität Ilmenau (TU Ilmenau), Germany. 2019 he was promoted to full professor after receiving the Heisenberg-professorship \textit{Optimization-based Control} by the German Research Foundation (DFG). He was recipient of the Ph.D.\ Award from the City of Bayreuth, Germany, and stipend of the German National Academic Foundation. 2013 he has been appointed Junior Fellow of the Society of Applied Mathematics and Mechanics (GAMM), where he served as speaker in 2014 and 2015. 
His current research interests include systems and control theory with a focus on nonlinear model predictive control, stability analysis, and data-driven techniques.

\begin{wrapfigure}[10]{l}{0.9in}
	\vspace{-8pt}
	\includegraphics[width=0.9in,height=1.15in,clip,keepaspectratio]{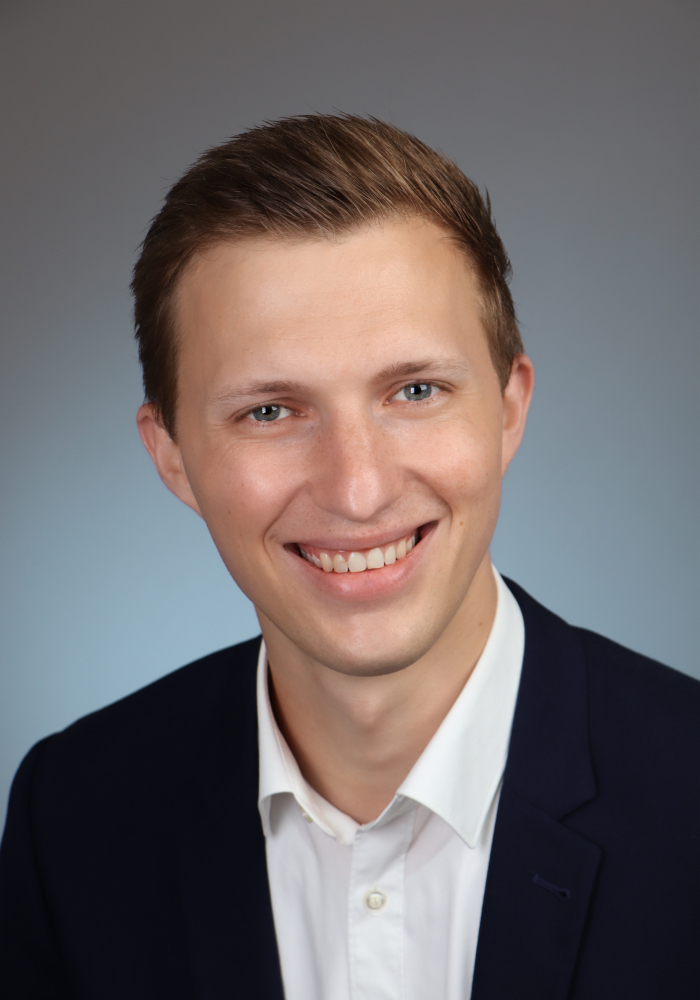}
	\vspace{-12pt}
\end{wrapfigure}
\noindent
\textbf{Julian Berberich} is a Lecturer (Akademischer Rat) at the Institute for Systems Theory and Automatic Control at the University of Stuttgart, Germany. He received his Ph.D. in Mechanical Engineering in 2022, and a Master’s degree in Engineering Cybernetics in 2018, both from the University of Stuttgart, Germany. In 2022, he was a visiting researcher at ETH Zürich, Switzerland. He is a recipient of the 2022 George S. Axelby Outstanding Paper Award as well as the Outstanding Student Paper Award at the 59th IEEE Conference on Decision and Control in 2020. His research interests include data-driven analysis and control as well as quantum computing.

\begin{wrapfigure}[10]{l}{0.9in}
	\vspace{-8pt}
	\includegraphics[width=0.9in,height=1.15in,clip,keepaspectratio]{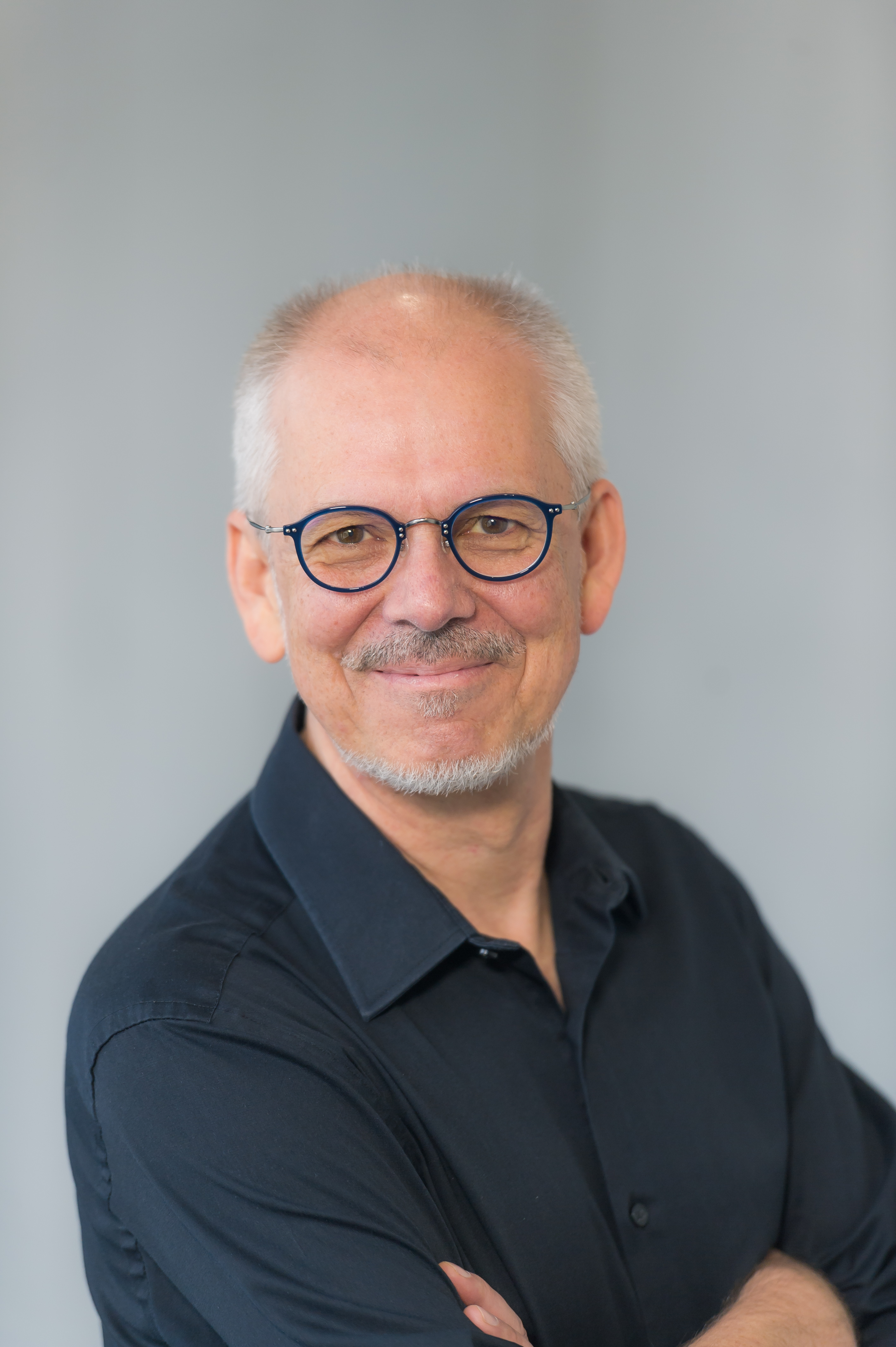}
	\vspace{-12pt}
\end{wrapfigure}
\noindent
\textbf{Frank Allg\"ower} studied engineering cybernetics and applied mathematics in Stuttgart and with the University of California, Los Angeles (UCLA), CA, USA, respectively, and received the Ph.D. degree from the University of Stuttgart, Stuttgart, Germany. Since 1999, he has been the Director of the Institute for Systems Theory and Automatic Control and a professor with the University of Stuttgart. His research interests include predictive control, data-based control, networked control, cooperative control, and nonlinear control with application to a wide range of fields including systems biology. Dr. Allgöwer was the President of the International Federation of Automatic Control (IFAC) in 2017–2020 and the Vice President of the German Research Foundation DFG in 2012–2020.

\egroup

\end{document}

%% file: sec1-introduction.tex
\section{Introduction}
Extended dynamic mode decomposition (EDMD; \cite{williams:kevrekidis:rowley:2015}) and its variants~\citep{colbrook:2024} are among the most popular machine learning algorithms to learn highly nonlinear dynamical (control) systems from data. 
Their applications range from climate forecasting~\citep{azencot:erichson:lin:mahoney:2020}, over electrocardiography~\citep{golany:radinstky:freedman:minha:2021} to quantum mechanics~\citep{klus:nuske:peitz:2022}, cryptography~\citep{strasser:schlor:allgower:2023}, and nonlinear fluid dynamics~\citep{mezic:2013}. 
The strength of EDMD lies in its analytical foundation by means of the Koopman operator, which \textit{propagates} observable functions (e.g., given by measurement sensors) along the flow of the underlying dynamical system. As the Koopman operator acts linearly on observables, EDMD leverages tools from regression to efficiently learn this linear object on a finite-dimensional set of observables, which renders the implementation of this learning algorithm relatively straightforward. 
Moreover, advanced tools from operator theory and linear regression enable the rigorous analysis of this machine learning algorithm \citep{kostic:novelli:maurer:ciliberto:rosasco:2022,philipp:schaller:worthmann:peitz:nuske:2024}.
For control tasks, in particular in safety-critical applications, such an analysis is key~\citep{folkestadt:chen:ames:burdick:2020}. 
The EDMD algorithm can be extended to control systems in order to take advantage of the reliability and strong analytical foundation. 
Two algorithms are predominantly used throughout the literature, namely EDMDc~\citep{korda:mezic:2018a,proctor:brunton:kutz:2018} and bilinear EDMD~\citep{williams:hemati:dawson:kevrekidis:rowley:2016,surana:2016}, where a linear and bilinear model is learned, respectively.
As such, EDMD with control has been extensively and successfully used in various applications, see~\cite{bevanda:sosnowski:hirche:2021} and the references therein.
\begin{figure*}[tb]
    \centering
    \scalebox{.8}{
    \begin{tikzpicture}[node distance=1.7cm,>=stealth]
        \node (dynSys) [rectangle, rounded corners, minimum width=7cm, minimum height=2.625cm,draw=black, fill=black!20,align=center,very thick]{\large\textsc{Dynamical System}\\[1ex]\large$\dot{\xb} = \fb(\xb) + \sum_{i=1}^m \gb_i(\xb)u_i$\\[0.75cm]};
        \node (DDsurrogate) [rectangle, rounded corners, minimum width=7cm, minimum height=2.625cm,draw=black, fill=black!20,align=center,very thick] at ($(dynSys.south east)+(4.125,-3.65)$) {~\\[1cm]\large\textsc{Data-Driven Bilinear}\\[1ex]\large\textsc{Surrogate Model}\\[1ex]\large$\Phib_+ \approx A \Phib + B_0 \ub + \sum_{i=1}^m u_iB_i\Phib$};
        \draw[->,very thick] ($(dynSys.south west)+(0.5,0)$) |- (DDsurrogate.west) node[pos=.75,below,rotate=0,align=center]{\large\textsc{Learning}\\[1ex]\large\textsc{via EDMD}};
        \draw[->,very thick] ($(DDsurrogate.north east)-(0.5,0)$) |- (dynSys.east) node[pos=.75,above,rotate=0,align=center] {\large\textsc{Controller with}\\[1ex]\large\textsc{Stability Guarantees}};
        \node [rectangle, rounded corners, minimum width=4cm, minimum height=4cm,draw=black, fill=white,align=center,anchor=north west] at ($(dynSys.south west)+(1.425,0.9)$) {
            \begin{tikzpicture}
                \begin{axis}[
                    domain=-1:1,
                    axis lines=none,
                    xtick=\empty, ytick=\empty,
                    clip mode=individual, clip=false,
                    xmin=-0.5, xmax=0.5,
                    ymin=-0.5, ymax=0.5,
                    width=3.5cm, height=3.5cm,
                ]  
                    \addplot [black!70, only marks, mark=*, samples=500, mark size=0.75]{rand};
                    \node[black,anchor=center,fill=black!20,minimum width=0cm,minimum height=0cm,fill opacity=0.9, text opacity=1] at (axis cs:0,0){\large\textsc{Data Samples}};
                \end{axis}
            \end{tikzpicture}
        };
        \node [rectangle, rounded corners, minimum width=4cm, minimum height=4cm,draw=black, fill=white,align=center,anchor=south east] at ($(DDsurrogate.north east)-(1.425,1.1)$) {
            \begin{tikzpicture}
                \begin{axis}[scale only axis,
                    axis lines=none,
                    xmin=-20,
                    xmax=20,
                    ymin=-21,
                    ymax=21,
                    width=6.72cm, height=4.56cm,
                    unbounded coords = jump,
                    restrict x to domain =-25:25,
                    restrict y to domain =-40:40,
                ]
                    \addplot[black!70,fill=black!70,smooth,mark=*] table [x index=4,y index=5] {data/exmp-cooked-up-eigenfunctions.dat};
                    \node[black,anchor=center,fill=black!20,minimum width=0cm,minimum height=0cm,fill opacity=0.9, text opacity=1,align=center] at (axis cs:0,4){\large\textsc{Region of}\\[1ex]\large\textsc{Attraction}};
                \end{axis}
            \end{tikzpicture}
        };
    \end{tikzpicture}}
    \caption{
        Illustration of the proposed controller design framework SafEDMD tailored to nonlinear data-driven control with closed-loop guarantees. 
        The proposed approach first uses EDMD to learn a bilinear surrogate model of the unknown nonlinear system based on data samples along with rigorous error bounds.
        Next, a controller is designed which exploits the model and the guaranteed error bounds to ensure closed-loop exponential stability of the nonlinear system in a guaranteed RoA.
    }
    \label{fig:sketch-highlevel}
\end{figure*}

The reliable control of complex systems based on EDMD requires a rigorous error analysis for the learning algorithm.
To this end,~\cite{korda:mezic:2018b} prove convergence of EDMD to the Koopman operator in the infinite-data limit in terms of their spectral properties, where~\cite{colbrook:mezic:stepanenko:2024} discuss fundamental limits.
Finite-data error bounds have been proven in~\cite{mezic:2022} and~\cite{zhang:zuazua:2023} for deterministic systems based on ergodic and i.i.d.\ sampling, respectively, and further refined in~\cite{philipp:schaller:boshoff:peitz:nuske:worthmann:2024} under less-restrictive assumptions, including discrete-time systems. 
The first extension to (stochastic) control systems is given in~\cite{nuske:peitz:philipp:schaller:worthmann:2023} including error bounds for bilinear EDMD. 
The error bounds in these works, however, are global in the sense that the error does not vanish at the origin, which prevents a straightforward application of standard control methods.

In this work, we present \textbf{S}tability- \textbf{a}nd \textbf{f}eedback-oriented \textbf{EDMD} (SafEDMD) --~a controller design framework with rigorous closed-loop guarantees for data-driven control of nonlinear systems.
The architecture relies on a structured EDMD algorithm to learn a bilinear surrogate model of a nonlinear system based on data samples.
To enable controller design with rigorous closed-loop guarantees, SafEDMD is accompanied by bounds on its approximation error, which, in contrast to existing results, vanish at the origin.
This allows us to design safe controllers, which we illustrate with the exemplary control objective of stabilization in a guaranteed region of attraction (RoA).
We note that~\cite{bold:grune:schaller:worthmann:2025} employs a related EDMD-based learning approach for the design of data-driven MPC with guaranteed practical asymptotic stability.
Further,~\cite{strasser:schaller:worthmann:berberich:allgower:2025} proposes a data-driven feedback design mechanism based on EDMD with closed-loop stability guarantees.
The key difference of the present paper to~\cite{bold:grune:schaller:worthmann:2025,strasser:schaller:worthmann:berberich:allgower:2025} is that the latter rely on sampling the Koopman \emph{generator}, requiring time-derivative data which are hard to obtain and are often very noisy in practice.
On the contrary, in the present paper, we employ the Koopman \emph{operator} directly in the discrete-time domain based on discrete input-state samples of the system.
This poses several technical challenges which are resolved via the proposed SafEDMD architecture and the corresponding feedback design.
To the best of our knowledge, the proposed approach is the first to control unknown nonlinear systems based on derivative-free data using the Koopman operator while giving rigorous closed-loop guarantees.

The overall approach is illustrated in Fig.~\ref{fig:sketch-highlevel} with the main contributions summarized as follows:
\begin{enumerate}
    \item [(1)] We propose SafEDMD --~a novel data-driven and highly efficient architecture for Koopman-based controller design with EDMD.
    \item [(2)] We derive proportional error bounds vanishing at the origin and tailored to controller design.
    \item [(3)] Based on the proposed EDMD-based bilinear surrogate and its error bounds, we design a feedback controller with closed-loop exponential stability guarantees.
    \item [(4)] We illustrate the methodology with various examples, highlighting scenarios where the proposed architecture allows us to successfully learn and control unknown nonlinear systems, whereas existing methods fail.
\end{enumerate}

The paper is organized as follows. 
In Section~\ref{sec:Koopman-operator}, we recall the notion of the Koopman operator for nonlinear dynamical systems.  
Section~\ref{sec:learning-architecture} describes the SafEDMD framework, introducing a Koopman-based bilinear surrogate model tailored to data-driven control.
Section~\ref{sec:controller-design} illustrates the SafEDMD controller design, where we design a data-driven feedback controller with closed-loop guarantees.
Finally, Section~\ref{sec:numerics} validates the proposed controller design framework in numerical experiments.

%% file: sec2-Koopman-operator.tex
\section{The Koopman operator as foundation for representing dynamical systems based on data}\label{sec:Koopman-operator}

In the following, we briefly recap the necessary background on the Koopman operator for nonlinear dynamical systems in Section~\ref{sec:Koopman-representation}. 
Then, in Section~\ref{sec:sampled-data-Koopman}, we discuss a sampled-data Koopman representation in preparation for the proposed data-driven controller design.

\subsection{Koopman representation of nonlinear dynamical systems}\label{sec:Koopman-representation}

Throughout the paper, we consider \emph{unknown} nonlinear control-affine systems of the form 
\begin{equation}\label{eq:dynamics-nonlinear}
    \dot{\xb}(t) = \fb(\xb(t)) + \sum_{i=1}^m\gb_i(\xb(t)) u_i(t),
\end{equation}
where $\xb(t) \in\bbR^n$ denotes the state at time $t \geq 0$ and the control function~$\ub: [0,\infty) \rightarrow \mathbb{R}^m$ serves as an input. 
The map $\fb:\bbR^n\to\bbR^n$ is called drift, while $\gb_i:\bbR^n\to\bbR^n$, $i \in [m]$, are called input maps\footnote{In this paper, we use $[a:b]\coloneqq\bbZ\cap[a,b]$ and $[b]\coloneqq [1:b]$.}. 
For an initial condition $\xb(0) = \hat{\xb}\in\bbR^n$ and a control function $\ub$, we denote the solution of~\eqref{eq:dynamics-nonlinear}, provided it exists, at time $t\geq 0$ by $\xb(t;\hat{\xb},\ub)$.
We assume that $\fb(\zerob)=\zerob$ holds, i.e., the origin is a controlled equilibrium for $\ub=\zerob$.

The Koopman operator $\cK_t^u$ introduced in~\cite{koopman:1931} provides a powerful alternative representation of the nonlinear dynamical system~\eqref{eq:dynamics-nonlinear} through the lens of observable functions given by, e.g., measurements in a particular application.
More precisely, the Koopman operator $\cK_t^{\ub}$ corresponding to~\eqref{eq:dynamics-nonlinear} with constant input $\ub(t) \equiv \ub \in \bbU$ for a compact set $\bbU\subseteq\bbR^m$ with $\zerob\in\operatorname{int}(\bbU)$ is defined as 
\begin{equation}\label{eq:Koopman-operator}
    (\cK_t^\ub \varphi)(\hat{\xb}) = \varphi(\xb(t;\hat{\xb},\ub))
\end{equation}
for all $t\geq 0$, $\hat{\xb}\in\bbX$, $\varphi \in L^2(\bbX,\bbR)$, where $\bbX\subseteq\bbR^n$ is a compact set and the real-valued functions $\varphi$ are called \emph{observables}.\footnote{We tacitly assumed invariance of $\bbX$ under the flow such that the observable functions are defined on $\xb(t;\hat{\xb},\ub)$ for all $\hat{\xb}\in \bbX$, which can be relaxed by considering initial values in a tightened version of $\bbX$, see~\cite{goor:mahony:schaller:worthmann:2023}.}
As the Koopman operator $\cK_t^{\ub}$ is an infinite-dimensional, but \emph{linear} operator on observable functions~$\varphi$, it enables the use of powerful data-driven techniques for learning such as, e.g., linear regression.

As shown, e.g., in~\cite{peitz:otto:rowley:2020,bold:grune:schaller:worthmann:2025,philipp:schaller:worthmann:peitz:nuske:2023b}, 
the Koopman operator $\cK_{t}^\ub$ inherits the control-affine structure of \eqref{eq:dynamics-nonlinear} \emph{approximately}, i.e., 
\begin{equation}\label{eq:Koopman-bilinear-approximate}
    \cK_t^\ub \approx \cK_t^\zerob + \sum_{i=1}^m u_i (\cK_t^{\eb_i}-\cK_t^\zerob)
\end{equation}
holds, where $\cK_t^\zerob$ and $\cK_t^{\eb_i}$, $i\in[m]$, are the Koopman operators corresponding to the constant control functions $\ub\equiv \zerob$ and $\ub\equiv \eb_i$ with unit vectors~$\eb_i$, $i\in[m]$, respectively. 
The resulting approximation error is precisely investigated in Theorem~\ref{thm:proportional-bound-data}.

The core of data-driven techniques leveraging the Koopman operator, such as EDMD as discussed later, is to learn the action of the Koopman operator on a subspace, called the \emph{dictionary}.
To this end, we define the dictionary $\bbV\coloneqq \spn{\phi_\ell}_{\ell=0}^{N}$ representing the $(N+1)$-dimensional subspace spanned by the chosen observables $\phi_0\equiv 1$ and $\phi_\ell\in\cC^1(\bbR^n,\bbR)$ with $\phi_\ell(\zerob)=0$, $\ell\in[N]$. 
In particular, we choose the observables
\begin{equation}\label{eq:lifting-function}
    \Phib(\xb) = \begin{bmatrix} 1 & \xb^\top & \phi_{n+1}(\xb) & \cdots & \phi_N(\xb) \end{bmatrix}^\top.
\end{equation}
Note that 
\begin{equation}\label{eq:philowerupper}
    \|\xb\|\leq\|\Phib(\xb)-\Phib(\zerob)\|\leq L_{\Phi}\|\xb\|,
\end{equation}
where the first inequality follows since $\Phib$ explicitly contains $\xb$, and the second inequality follows from local Lipschitz continuity due to $\Phib\in\cC^1(\bbR^n,\bbR^N)$.
Common choices for the observables $\Phib$ include monomials or radial basis functions such as, e.g., thin-plate splines. 
If prior knowledge about the system dynamics is available, suitable observables can often be inferred from the system dynamics, see, e.g.,~\cite{haseli:cortes:2023c} or~\cite{shi:karydis:2021} for an approach to construct a dictionary tailored to robotic systems.
Although not explicitly considered in this paper, the observables $\phi_\ell$, $\ell\in[n+1,N]$, can also be learned via a neural network (NN). 
Parametrizing $\Phib$ by an NN increases the expressiveness and allows finding suitable nonlinear observables $\phi_\ell$ for the underlying system dynamics~\eqref{eq:dynamics-nonlinear}, see, e.g.,~\cite{takeishi:kawahara:yairi:2017,yeung:kundu:hodas:2019,shi:meng:2022}.
Here, the assumed Lipschitz continuity in~\eqref{eq:philowerupper} of the learned NN can be ensured by design~\citep{pauli:koch:berberich:kohler:allgower:2021}.

In~\eqref{eq:lifting-function}, we assume that the full state is contained in the dictionary. Preliminary results, however, motivate that dictionaries being observable from a system-theoretic point of view may suffice to design a stabilizing controller, where the state coordinates are reconstructed via, e.g., delay coordinates~\citep{otto2024learning}. A rigorous investigation of non-state-inclusive dictionaries is left for future work.

\subsection{Sampled-data Koopman representation}\label{sec:sampled-data-Koopman}

When controlling dynamical systems of the form~\eqref{eq:dynamics-nonlinear}, a direct implementation of a (piecewise) continuous control function~$\ub: [0,\infty) \rightarrow \mathbb{R}^m$ is typically not feasible.
Instead, a common strategy is to sample the system~\eqref{eq:dynamics-nonlinear}, e.g., equidistantly in time, and consider piecewise constant control inputs on each interval, i.e., $\ub(t)\equiv \ub_k\in\bbU$ holds for all $t \in [t_k,t_k+\Delta t)$ with sampling period $\Delta t > 0$ and $t_k=k\Delta t$, $k \in \mathbb{N}_0$.
This leads to the discretized system representation
\begin{equation}\label{eq:dynamics-nonlinear-sampled}
    \xb_{k+1} 
    = \xb_k + \int_{t_k}^{t_k+\Delta t} \fb(\xb(t)) + \sum_{i=1}^m \gb_i(\xb(t))(\ub_{k})_i\,\mathrm{d}t
\end{equation}
with $\xb(t) = \xb(t;\hat{\xb},\ub)$ and $\xb_k = \xb(t_k;\hat{\xb},\ub)$.
From~\eqref{eq:Koopman-operator}, we deduce the discrete-time Koopman representation
\begin{equation}\label{eq:lifted-dynamics-general}
    \Phib(\xb_{k+1}) = (\cK_{\Delta t}^\ub \Phib)(\xb_k).
\end{equation}
Note that the evolution of $\Phib(\xb_k)$ allows to retrieve the underlying state $\xb$ by projection due to the choice of the lifting function $\Phib$ in~\eqref{eq:lifting-function}, such that $\xb=\begin{bmatrix} \zerob_{n\times 1} & I_n & 0_{n\times N-n}\end{bmatrix}\Phib(x)$ by definition. 

%% file: sec3-learning-architecture.tex
\section{SafEDMD: a Koopman-based bilinear surrogate model of nonlinear dynamical systems}\label{sec:learning-architecture}

In the following, we present a Koopman-based bilinear surrogate model tailored to the controller design for nonlinear systems with closed-loop guarantees. 
Since the true nonlinear system dynamics are unknown, we leverage the linearity of the Koopman operator in the observables to learn a data-driven approximation with rigorous error bounds.
More precisely, we introduce a novel architecture called \emph{SafEDMD}.
This framework 1) leverages a structured EDMD variant to learn a data-based bilinear surrogate model of the Koopman operator solely relying on linear regression and 2) ensures error bounds that allow the usage of the surrogate model for reliable control tasks, e.g., closed-loop stabilization.

To determine data-driven estimates of $\cK_{\Delta t}^\ub$ for constant control inputs $\ub(t)\equiv\bar{\ub}$ for $\bar{\ub}\in\{\zerob,\eb_1,...,\eb_m\}$, we consider data samples $\cD=\{\xb^{\bar{\ub}}_j,\yb^{\bar{\ub}}_j\}_{j=1}^{d^{\bar{\ub}}}$, where $\yb^{\bar{\ub}}_j=\xb(\Delta t; \xb^{\bar{\ub}}_j,\bar{\ub})$ and the control inputs $\eb_i$, $i\in[m]$, are the unit vectors.
In particular, we need no information about the state derivative but only require trajectory samples of the state and its successor of an unknown nonlinear system~\eqref{eq:dynamics-nonlinear} for a set of chosen control values. 
Note that we choose the canonical unit vectors as control values to collect $m+1$ samples at each (state) data point, but any other basis of $\bbR^m$ can be chosen in addition to the zero input.
Importantly, the chosen sampling allows us to exploit the approximately control-affine structure~\eqref{eq:Koopman-bilinear-approximate} of the Koopman operator for the \emph{controlled} system, which enables the investigation of each (autonomous) Koopman operator $\cK_t^{\bar{\ub}}$ separately.
We refer to~\cite{bevanda:driessen:iacob:toth:sosnowski:hirche:2024,bold:philipp:schaller:worthmann:2024,strasser:schaller:berberich:worthmann:allgower:2025} for more flexible sampling schemes w.r.t.\ the control input $u$ and defer their use within our controller design framework to future work.

The proposed learning approach exploits the fact that the Koopman operator acting on the defined lifting function $\Phib$ has a specific structure.
In particular, note that $\Phib$ contains a constant observable $\phi_0\equiv 1$, i.e., $\phi_0(\xb_{k+1})=\phi_0(\xb_k)=1$, and, thus, $\Phib(\zerob)=(\cK^\zerob_{\Delta t}\Phib)(\zerob)$ due to $\fb(\zerob)=\zerob$.
Therefore, using~\eqref{eq:Koopman-bilinear-approximate}, we partition the function space into constant functions and the remainder to obtain
\begin{subequations}\label{eq:structure-Koopman-operator}
    \begin{gather}
        \cK_{\Delta t}^{\ub} = \begin{bmatrix}
            1 & \zerob^\top \\ (\cK_{\Delta t}^{\ub})_{21} & (\cK_{\Delta t}^{\ub})_{22}
        \end{bmatrix}
        ,
        \\
        \cK_{\Delta t}^\zerob = \begin{bmatrix}
            1 & \zerob^\top \\ \zerob & (\cK_{\Delta t}^\zerob)_{22}
        \end{bmatrix}    
        ,\;
        \cK_{\Delta t}^{\eb_i} = \begin{bmatrix}
            1 & \zerob^\top \\ (\cK_{\Delta t}^{\eb_i})_{21} & (\cK_{\Delta t}^{\eb_i})_{22}
        \end{bmatrix}
    \end{gather}
\end{subequations}
as the corresponding partition of the Koopman operators.
Due to the structure of the Koopman operator in~\eqref{eq:structure-Koopman-operator}, we enforce the same structure for the data-driven surrogates in our proposed architecture, i.e., we adopt the structure in the finite-dimensional model given by
\begin{equation}\label{eq:structure-Koopman-operator-data}
    \cK_{\Delta t,d}^\zerob 
    = \begin{bmatrix} 
        1 & \zerob^\top \\ \zerob & A
    \end{bmatrix} 
    ,\quad
    \cK_{\Delta t,d}^{\eb_i} 
    = \begin{bmatrix} 
        1 & \zerob^\top \\ \bb_{0,i} & \hB_i
    \end{bmatrix}.
\end{equation}
To learn the unknown matrices $A$, $\bb_{0,i}$, $\hB_i$, we arrange the data~$\cD$ in
\begin{subequations}\label{eq:data-arrangement}
    \begin{align}        
        X^\zerob
        &= \begin{bmatrix}
            \zerob_{N\times 1} & I_N
        \end{bmatrix}
        \begin{bmatrix}
            \Phib(\xb^\zerob_1)
            & \cdots & 
            \Phib(\xb^\zerob_{d^\zerob})
        \end{bmatrix},
        \\
        X^{\eb_i}
        &= \begin{bmatrix}
            \Phib(\xb^{\eb_i}_1) & \cdots & \Phib(\xb^{\eb_i}_{d^{\eb_i}})
        \end{bmatrix},
        \\
        Y^{\bar{\ub}}
        &= \begin{bmatrix}
            \zerob_{N\times 1} & I_N
        \end{bmatrix}
        \begin{bmatrix}
            \Phib(\yb^{\bar{\ub}}_1)
            & \cdots & 
            \Phi(\yb^{\bar{\ub}}_{d^{\bar{\ub}}})
        \end{bmatrix}.
    \end{align}
\end{subequations}
Then, as in EDMD, the proposed architecture relies on solving the linear regression problems
\begin{subequations}\label{eq:EDMD}
    \begin{align}
        A
        &= \argmin_{A\in\bbR^{N\times N}} 
        \| Y^\zerob - A X^\zerob \|_\mathrm{F},
        \\
        \begin{bmatrix} 
            \bb_{0,i} & \hB_i
        \end{bmatrix}
        &= \argmin_{\substack{\bb_{0,i}\in\bbR^{N},\\\hB_i\in\bbR^{N\times N}}}
        \|Y^{\eb_i} - \begin{bmatrix} \bb_{0,i} & \hB_i\end{bmatrix} X^{\eb_i}\|_\mathrm{F}
    \end{align}
\end{subequations}
for $i\in[m]$, where $\|\cdot\|_\mathrm{F}$ denotes the Frobenius norm.
Based on the estimates in~\eqref{eq:EDMD}, we define $
    B_0=\left[\begin{smallmatrix}
        \bb_{0,1} & \cdots & \bb_{0,m}
    \end{smallmatrix}\right]
$ and $
    \tB = \left[\begin{smallmatrix}
        \hB_1 - A & \cdots & \hB_m - A
    \end{smallmatrix}\right]
$. 
Further, we can construct a data-driven surrogate model for the discretization of the nonlinear dynamical system~\eqref{eq:dynamics-nonlinear}, i.e., 
$
    \cK_{\Delta t,d}^\ub=\cK_{\Delta t,d}^\zerob + \sum_{i=1}^m u_i (\cK_{\Delta t,d}^{\eb_i}-\cK_{\Delta t,d}^\zerob).
$
This model, however, is only an approximation of the true Koopman operator in~\eqref{eq:lifted-dynamics-general}, meaning that
\begin{equation}\label{eq:dynamics-operator-surrogate-approx}
    \Phib(\xb_{k+1})\approx\cK_{\Delta t,d}^\ub\Phib(\xb_k).
\end{equation}
The error is due to the different sources, including the bilinear approximation~\eqref{eq:Koopman-bilinear-approximate} as well as the above data-driven estimation.
In the following main result, which forms the foundation of the proposed control architecture, we provide a rigorous analysis of this error.
\begin{theorem}\label{thm:proportional-bound-data}
    Suppose that the data samples~$\cD$ are i.i.d. 
    Then, for any probabilistic tolerance $\delta \in (0, 1)$, amount of data $d_0 \in \bbN$ and sampling rate $\Delta t>0$, there are constants $\bar{c}_x,\bar{c}_u = \mathcal{O}\left(\nicefrac{1}{\sqrt{\delta d_0}} + \Delta t^2\right)$ such that for all $d \geq d_0$, the learning error bound\footnote{$P_\bbV$ denotes the $L^2$-orthogonal projection onto $\bbV$.}
    \begin{multline}\label{eq:proportional-bound-data}
         \|(P_\bbV\cK_{\Delta t}^\ub|_\bbV)\Phib(\xb)-\cK_{\Delta t,d}^\ub\Phib(\xb)\|
         \\
         \leq \bar{c}_x \|\Phib(\xb)-\Phib(\zerob)\| + \bar{c}_u \|\ub\|
    \end{multline}
    holds for all $\xb \in \bbX$ and $\ub \in \bbU$ with probability $1-\delta$.
\end{theorem}

\textbf{PROOF} (Sketch)\textbf{.}\;
    We prove the result in two steps, see Appendix~\ref{app:proof-proportional-bound-DD-surrogate} for details. First, we bound the error of the bilinear representation for the projected Koopman operator $P_\bbV \cK_{\Delta t}^\ub|_\bbV$ in~\eqref{eq:Koopman-bilinear-approximate}. Then, we incorporate the estimation error from learning the bilinearized Koopman operator from data.
\null\hfill$\square$

The error bound~\eqref{eq:proportional-bound-data} of Theorem~\ref{thm:proportional-bound-data} features a central ingredient crucial for design that is not present in existing estimates, as recalled in Proposition~\ref{prop:error-bound-operator} in Appendix~\ref{sec:data-driven-surrogate-Koopman}.
In particular,~\cite{nuske:peitz:philipp:schaller:worthmann:2023} derives a bound on the estimation error which does not vanish at the origin. 
Thus, this bound cannot be used within a robust control setting to guarantee asymptotic (or exponential) stability of the closed-loop system.
To address this issue, Theorem~\ref{thm:proportional-bound-data} provides a tailored error bound suitable for control.
First, this bound is proportional in the sense that the right-hand side vanishes for $(\xb,\ub)=\zerob$. 
Such a proportional bound is, to the best of the authors' knowledge, so far not present in any learning scheme using the notably simple EDMD framework and is decisive for controller design such that the underlying true system is reliably and safely operated by feedback control in a region containing the origin. 
Second, the proportionality constants $\bar{c}_x$ and $\bar{c}_u$ in~\eqref{eq:proportional-bound-data} can be made arbitrarily small when choosing a sufficiently high amount of data points $d_0$ and a small enough sampling rate $\Delta t$.
This is important for a successful controller design since the considered closed-loop objective requires robustness against all possible learning errors satisfying the bound in~\eqref{eq:proportional-bound-data}. 
While we consider i.i.d. sampling in this paper, future work should investigate proportional error bounds for other sampling strategies (e.g.~ergodic sampling), where we conjecture that similar arguments as in Theorem~\ref{thm:proportional-bound-data} apply if the estimation error of the data-based Koopman estimate $\|P_\bbV\cK_{\Delta t}^\ub|_\bbV - K_{\Delta t,d}^\ub\|$ is bounded.
Further, motivated by the experiments conducted in Section~\ref{sec:numerics}, we expect that the result in Theorem~\ref{thm:proportional-bound-data} generalizes to the case of noisy data, see, e.g.,~\cite{llamazares:llamazares:latz:klus:2024,chatzikiriakos:strasser:allgower:iannelli:2024} for related bounds on the Koopman estimate.

The following corollary provides a proportional bound on the complete approximation error, where an additional projection error due to a \emph{finite} dictionary of observables is taken into account.
\begin{corollary}\label{cor:proportional-bound-error}
    Let the assumptions of Theorem~\ref{thm:proportional-bound-data} hold. 
    If there exists a proportional error bound on the projection error, i.e., 
    \begin{multline}\label{eq:proportional-bound-projection}
        \|(\cK_{\Delta t}^\ub \Phib)(\xb) - (P_\bbV\cK_{\Delta t}^\ub|_\bbV)\Phib(\xb) \| 
        \\
        \leq \tilde{c}_x \|\Phib(\xb)-\Phib(\zerob)\| + \tilde{c}_u \|\ub\|,
    \end{multline}
    then for any probabilistic tolerance $\delta \in (0, 1)$, amount of data $d_0 \in \bbN$ and sampling rate $\Delta t>0$, there are constants $\bar{c}_x,\bar{c}_u = \mathcal{O}\left(\nicefrac{1}{\sqrt{\delta d_0}} + \Delta t^2\right)$ such that for all $d \geq d_0$, the full approximation error $\|(\cK_{\Delta t}^\ub \Phib)(\xb) - \cK_{\Delta t,d}^\ub\Phib(\xb)\|$ is proportionally bounded by
    \begin{equation}\label{eq:proportional-bound-error}
        \|(\cK_{\Delta t}^\ub \Phib)(\xb) - \cK_{\Delta t,d}^\ub\Phib(\xb) \| 
        \leq c_x \|\Phib(\xb)-\Phib(\zerob)\| + c_u \|\ub\|,
    \end{equation}
    for all $\xb\in\bbX$ and $\ub\in\bbU$ with probability $1-\delta$, where $c_x = \bar{c}_x+\tilde{c}_x$ and $c_u=\bar{c}_u+\tilde{c}_u$.
\end{corollary}
\begin{pf}
 As a direct consequence of Theorem~\ref{thm:proportional-bound-data}, we compute
 \begin{align*}
        &\|(\cK_{\Delta t}^\ub \Phib)(\xb) - \cK_{\Delta t,d}^\ub\Phib(\xb)\|
        \\
        &\quad\leq \|(\cK_{\Delta t}^\ub \Phib)(\xb) - (P_\bbV\cK_{\Delta t}^\ub|_\bbV)\Phib(\xb) \| 
        \\
        &\quad\qquad\qquad\qquad
        + \|\left((P_\bbV\cK_{\Delta t}^\ub|_\bbV)-\cK_{\Delta t,d}^\ub\right)\Phib(\xb)\|
        \\
        &\quad\leq (\bar{c}_x + \tilde{c}_x) \|\Phib(\xb)-\Phib(\zerob)\| + (\bar{c}_u + \tilde{c}_u) \|\ub\|
    \end{align*}
    using the triangle inequality.
\end{pf}
The proportional bound~\eqref{eq:proportional-bound-projection} trivially holds if, e.g., the dictionary $\bbV$ is invariant w.r.t. the dynamics, i.e., $P_\bbV \cK^\ub_t|_\bbV = \cK^\ub_t|_\bbV$. 
This property is commonly employed in Koopman-based control~\citep{brunton:brunton:proctor:kutz:2016,goswami:paley:2017,huang:ma:vaidya:2018,mauroy:sootla:mezic:2020,goswami:paley:2020,goswami:paley:2021,schulze:doncevic:mitsos:2022}. 
Conditions for the (approximate) satisfaction of this invariance are given by, e.g.,~\cite{brunton:brunton:proctor:kutz:2016,korda:mezic:2020,goswami:paley:2021,haseli:cortes:2023c}. 
In this case, one has to additionally account for the 
projection error~\citep{schaller:worthmann:philipp:peitz:nuske:2023}. 
In~\cite[Sec.~4]{iacob:toth:schoukens:2024}, the authors suggest to derive uniform bounds based on polynomial tests, where interpolation arguments may be used to ensure the proportional bound as in~\eqref{eq:proportional-bound-projection} on the projection error, see, e.g.,~\cite{yadav:mauroy:2025,kohne:philipp:schaller:schiela:worthmann:2025} for recently derived uniform error bounds w.r.t.\ autonomous systems and~\cite{strasser:schaller:berberich:worthmann:allgower:2025} for the respective extension to bilinear surrogates for controlled systems.

The proposed learning architecture enabling data-driven control of nonlinear dynamical systems based on the linear regression problems~\eqref{eq:EDMD} and the error bound~\eqref{eq:proportional-bound-error} is summarized in Algorithm~\ref{alg:learning-architecture}.
Here, we emphasize that the EDMD-based identification is similar to schemes proposed in the literature, see, e.g.,~\cite{peitz:otto:rowley:2020}, but Corollary~\ref{cor:proportional-bound-error} is the first rigorous characterization of the resulting approximation error from finite data. 
This is exploited in the following to design a controller with rigorous closed-loop stability guarantees for the underlying nonlinear system.
\begin{algorithm}[tb]
   \caption{SafEDMD: A Koopman-based data-driven representation of nonlinear dynamical systems}
   \label{alg:learning-architecture}
    \begin{algorithmic}
       \STATE {\bfseries Input:} data $\cD = \{\xb_j^{\bar{\ub}},\yb_j^{\bar{\ub}}\}_{j=1}^d$ for $\bar{\ub}=\zerob,\eb_1,...,\eb_m$ with $\yb_j^{\bar{\ub}}=\xb(\Delta t; \xb^{\bar{\ub}}_j,\bar{\ub})$,
       observables $\Phib$
       \STATE Arrange data in~\eqref{eq:data-arrangement}.
       \STATE Solve the EDMD problem~\eqref{eq:EDMD}.
       \STATE {\bfseries Output:} Koopman representation~\eqref{eq:dynamics-operator-surrogate-approx} with $\cK^\zerob_{\Delta t,d}$, $\cK^{\eb_1}_{\Delta t,d},...,\cK^{\eb_m}_{\Delta t,d}$ in~\eqref{eq:structure-Koopman-operator-data} satisfying the error bound~\eqref{eq:proportional-bound-error}
    \end{algorithmic}
\end{algorithm}

%% file: sec4-controller-design.tex
\section{Data-driven controller design for nonlinear systems with closed-loop guarantees}\label{sec:controller-design}
\begin{figure}[tb]
    \vskip -0.01in
    \centering
    \resizebox{0.475\textwidth}{!}{
        \pgfdeclarelayer{background}
        \pgfsetlayers{background,main}
        \begin{tikzpicture}[node distance=1.7cm,>=stealth]
            \node (start) [rectangle,rounded corners,minimum height=1cm,text centered,draw=black,minimum width=6.91cm,align=center,fill=black!20] {\large$\dot{\xb} = \fb(\xb) + \sum_{i=1}^m \gb_i(\xb)u_i$\\\large with \textsc{data samples}};
            \node (lift) [rectangle,rounded corners,minimum height=1cm,text centered,draw=black,fill=black!10,anchor=north east,minimum width=5.1cm] at ($(start.south west) + (9cm,-1.2)$) {\large$\Phib_+ = \mathcal{K}^\ub_{\Delta t}\Phib$};
            \draw [->,very thick] ($(start.south east) + (-0.5,0)$) -- node[anchor=east,xshift=-.2cm,pos=0.75] (liftText) {\large\textsc{lift}} (lift);
            \node (bilin) [rectangle,rounded corners,minimum width=3cm, minimum height=1cm,text centered, draw=black,fill=black!10,anchor=north east] at ($(lift.south east) + (0,-0.75)$) {\large$\Phib_+ \approx \left(\mathcal{K}^\zerob_{\Delta t} + \sum_{i=1}^m u_i \left(\mathcal{K}^{\eb_i}_{\Delta t} - \mathcal{K}^{\zerob}_{\Delta t}\right)\right)\Phib$};
            \draw [->,very thick] (lift) -- node[anchor=east,xshift=-.2cm] {\large\textsc{bilinearize}} ($(lift.south) + (0,-0.75)$);
            \node (data) [rectangle,rounded corners,minimum height=1cm,text centered,draw=black,fill=black!10,anchor=north east,minimum width=9cm] at ($(bilin.south east) + (0,-1)$) {\large$\Phib_+ \approx \left( \mathcal{K}^\zerob_{\Delta t,d} + \sum_{i=1}^m u_i \left( \mathcal{K}^{\eb_i}_{\Delta t,d} - \mathcal{K}^{\zerob}_{\Delta t,d}\right)\right)\Phib$};
            \draw [-,very thick] let \p1 = (lift), \p2 = (bilin.south), \p3 = (data.north) in (\x1,\y2) |- ($(data.north) + (0,0.4)$);
            \draw [->,very thick] ($(start.south west)+ (0.5,0)$) |- ($(data.north) + (0,0.4)$) node[above] {\large\textsc{learn}} -- (data.north);
            \draw [->,very thick] (data.east) -- +(1,0) |- node[pos=0.45,anchor=east,text width=2.25cm,xshift=0.1cm] {\large Feedback $\ub=\mub(\xb)$} (start);
            \begin{pgfonlayer}{background}
                \node[rectangle,rounded corners,draw=black,fit=(lift) (bilin) (data) (liftText),align=left,fill=black!20] (safEDMDbox){};
            \end{pgfonlayer} 
            \node[anchor=south,rotate=90] (SafEDMD) at ($(safEDMDbox.west) + (0,0)$){\large\textsc{SafEDMD}};
        \end{tikzpicture}
    }
    \vspace*{-1.5\baselineskip}
    \caption{Data-driven controller design framework for nonlinear dynamical systems with closed-loop guarantees.
        The overall scheme consists of two main steps:
        1) learning a bilinear surrogate model along with error bounds; 
        2) designing a feedback controller $\ub=\mub(\xb)$ which robustly stabilizes the surrogate model.
        The approach guarantees closed-loop exponential stability due to the rigorous foundation based on SafEDMD, which learns a bilinear approximation of the Koopman operator dynamics and corresponding error bounds using data.
    }
    \label{fig:sketch-controller}
\end{figure}
In this section, we use the SafEDMD-based bilinear surrogate model derived in Section~\ref{sec:learning-architecture} to design a stabilizing controller for System~\eqref{eq:dynamics-nonlinear}. 
More precisely, our goal is to find a state-feedback controller $\mub:\bbR^n\to\bbR^m$ exponentially stabilizing the setpoint in closed loop using the control input $\ub_k=\mub(\xb_k)$.
Without loss of generality, we assume that the setpoint is the origin $\xb=\zerob$.
To find a suitable controller, we first proceed as in Algorithm~\ref{alg:learning-architecture} to obtain a data-driven surrogate model of the unknown nonlinear system based on sampled data.
We then use this model along with the error bound in Corollary~\ref{cor:proportional-bound-error} to design a data-driven controller with closed-loop stability guarantees.
Here, it is sufficient to bound the residual error in the lifted space since we aim for a \emph{robust} controller based on the bilinear surrogate model.
The overall control scheme is illustrated in Fig.~\ref{fig:sketch-controller}.

To obtain rigorous stability guarantees, our proposed controller ensures that the system trajectories evolve in a predetermined set $\mathbf{\Delta}_{\Phib}=\{\psib\in\bbR^N\mid\eqref{eq:DeltaPhi-definition}~\text{holds}\}$ with 
\begin{equation}\label{eq:DeltaPhi-definition}
    \begin{bmatrix}
        \psib \\ 1
    \end{bmatrix}^\top 
    \begin{bmatrix}
        Q_z & \sb_z \\ \sb_z^\top & r_z
    \end{bmatrix}
    \begin{bmatrix}
        \psib \\ 1
    \end{bmatrix}
    \geq 0
\end{equation}
for fixed matrices $Q_z\in\bbR^{N\times N}$, $\sb_z\in\bbR^N$, $r_z\in\bbR$ with $Q_z\prec 0$ and $r_z>0$ for which the inverse 
$
    \left[\begin{smallmatrix}
        \tQ_z & \tsb_z \\ \tsb_z^\top & \tr_z
    \end{smallmatrix}\right]
    \coloneqq 
    \left[\begin{smallmatrix}
        Q_z & \sb_z \\ \sb_z^\top & r_z
    \end{smallmatrix}\right]^{-1}
$
exists.
In particular, we enforce $\hat{\Phib}(\xb)\in\mathbf{\Delta}_{\Phib}$ for all times, where $\hat{\Phib}\coloneqq\begin{bmatrix}0_{N\times 1} & I_N\end{bmatrix}\Phib(\xb)$ denotes the reduced observable function after removing the constant observable $\phi_0\equiv 1$ such that $\hat{\Phib}(\zerob)=\zerob$.
The parametrization of $\mathbf{\Delta}_{\Phib}$ includes, e.g., a simple norm bound $\|\hat{\Phib}(\xb)\|^2\leq c$ for all $\xb$ by choosing $Q_z=-I$, $\sb_z=0$, and $r_z=c$.
The set $\mathbf{\Delta}_{\Phib}$ needs to be defined by the user before applying the following controller design procedure.
Ideally, one chooses $\mathbf{\Delta}_{\Phib}$ preferably large since it determines an outer bound on the RoA in which the proposed controller reliably controls the system.
However, there is a trade-off since too large choices of $\mathbf{\Delta}_{\Phib}$ may lead to an infeasible controller design.
In Section~\ref{sec:numerics}, we discuss practical possibilities for choosing $\mathbf{\Delta}_{\Phib}$ using numerical examples.

The following theorem yields a controller-design method guaranteeing closed-loop exponential stability of the \emph{sampled} nonlinear system~\eqref{eq:dynamics-nonlinear-sampled} by solving a linear matrix inequality feasibility problem.
\begin{theorem}\label{thm:stability-condition-LFR}
    Let the data points $\cD$ be sampled i.i.d.~from $\bbX$. 
    Suppose a data-driven surrogate model of the unknown nonlinear system~\eqref{eq:dynamics-nonlinear} obtained via the learning architecture in Algorithm~\ref{alg:learning-architecture} satisfying the error bound~\eqref{eq:proportional-bound-error} with given probabilistic tolerance $\delta \in (0,1)$, sampling rate $\Delta t > 0$, and constants $c_x,c_u = \mathcal{O}\left(\nicefrac{1}{\sqrt{\delta d_0}} + \Delta t^2\right)$.
    \\
    Further, assume there is a matrix $0\prec P=P^\top\in \bbR^{N\times N}$, matrices $L\in\bbR^{m\times N}$, $L_w\in\bbR^{m\times Nm}$, a matrix $0\prec \Lambda=\Lambda^\top \in \bbR^{m\times m}$, and scalars $\nu>0$, $\tau>0$
    such that~\eqref{eq:stability-condition-LFR} at the top of the next page and%
    \begin{figure*}[bt]
        \begin{equation}\label{eq:stability-condition-LFR}
           \hspace*{-0.01\textwidth}
            \begin{bmatrix}
                P - \tau I_{N}
                & - \tB (\Lambda\kron \tsb_z) - B_0 L_w(I_m\kron\tsb_z)
                & 0
                & AP + B_0L
                & \tB(\Lambda \kron I_N) + B_0L_w
                \\ 
                \star
                & (\Lambda \kron \tr_z) - L_w(I_m\kron\tsb_z) - (I_m\kron\tsb_z^\top)L_w^\top
                & -(I_m\kron\tsb_z^\top)\left[\begin{smallmatrix}0 \\ L_w\end{smallmatrix}\right]^\top
                & L
                & L_w
                \\
                \star
                & \star
                & 0.5\tau \left[\begin{smallmatrix} c_x^{-2} I_N & 0 \\ 0 & c_u^{-2}I_m \end{smallmatrix}\right]
                & \left[\begin{smallmatrix}P\\L\end{smallmatrix}\right]
                & -\left[\begin{smallmatrix}0 \\ L_w\end{smallmatrix}\right]
                \\
                \star 
                & \star
                & \star
                & P
                & 0
                \\
                \star
                & \star
                & \star
                & \star
                & - (\Lambda \kron \tQ_z^{-1})
            \end{bmatrix}
            \succ 0,
        \end{equation}
        \hrule
    \end{figure*}
    \begin{equation}\label{eq:stability-condition-LFR-invariance}
        \begin{bmatrix}
            \nu\tr_z - 1 & -\nu\tsb_z^\top \\ 
            -\nu\tsb_z & \nu\tQ_z + P 
        \end{bmatrix}
        \preceq 0
    \end{equation}
    hold.
    Then there exists an amount of data $d_0\in\bbN$ such that for all $d\geq d_0$ the controller\footnote{By $\kron$ we refer to the Kronecker product.}
    \begin{equation}\label{eq:controller-parametrization-nonlinear}
        \mub(\xb) = (I-L_w(\Lambda^{-1}\kron\hat{\Phib}(\xb)))^{-1} L P^{-1} \hat{\Phib}(\xb)
    \end{equation}
    ensures exponential stability of the \emph{sampled} nonlinear system~\eqref{eq:dynamics-nonlinear-sampled} for all initial conditions in the RoA $\hat{\xb}\in\cX_\mathrm{RoA} := \{\xb\in\bbR^n\,|\, \hat{\Phib}(\xb)^\top P^{-1} \hat{\Phib}(\xb) \leq 1\}$ with probability $1-\delta$.
\end{theorem}

\textbf{PROOF} (Sketch)\textbf{.}\;
    The result is shown in two steps. 
    First, we show that all $\xb\in\cX_\mathrm{RoA}$ satisfy $\hat{\Phib}(\xb)\in\mathbf{\Delta}_\Phi$.
    Then, we conclude positive invariance of $\cX_\mathrm{RoA}$ together with exponential stability of the sampled closed-loop system~\eqref{eq:dynamics-nonlinear-sampled} for all $\hat{\xb}\in\cX_\mathrm{RoA}$.
    This ensures closed-loop exponential stability under the obtained feedback via the error bounds of the learning architecture in Algorithm~\ref{alg:learning-architecture}, see Appendix~\ref{app:proof-thm-stability-condition-LFR} for details.
\null\hfill$\square$

The proposed controller design in Theorem~\ref{thm:stability-condition-LFR} is purely based on data samples of the state trajectory and does not rely on model knowledge.
In particular, we use the SafEDMD architecture satisfying the error bound specified in Corollary~\ref{cor:proportional-bound-error} to obtain a reliable data-driven system parametrization. 
Learning the Koopman-based surrogate model via Algorithm~\ref{alg:learning-architecture} only requires solving the least-squares optimization problem~\eqref{eq:EDMD}, which has complexity $\cO(N^3)$ and scales well to higher-dimensional systems~\citep{klus:bittracher:schuster:schutte:2018,nuske:klus:2023}. 
The controller design approach, on the other hand, is framed as a semi-definite program using linear matrix inequalities with complexity $\cO(N^6)$. 
Although efficiently solvable for a moderate dictionary size $N$~\citep{vandenberghe:boyd:1996}, future work should be devoted to including structure-exploiting SDP techniques, see, e.g.,~\cite{deklerk:2010,gramlich:holicki:scherer:ebenbauer:2023}.
In Section~\ref{sec:numerics}, we show that the proposed approach can successfully control common benchmark systems in Koopman theory with more reliable behavior than existing approaches.
The robust controller uses the learning error bounds in~\eqref{eq:proportional-bound-error} to ensure closed-loop \emph{exponential} stability.
In particular, compared to other results ensuring only \emph{practical} stability, we obtain exponential stability due to the established novel error bounds of Corollary~\ref{cor:proportional-bound-error} that vanish at the equilibrium.
Here, we can directly relate a given desired accuracy and certainty about the closed-loop guarantees as well as the input dimension $m$ to the necessary amount of data $d_0$ and the sampling period $\Delta t$.

We note that the controller parametrization~\eqref{eq:controller-parametrization-nonlinear} contains the control law $\mub(\xb)=K\hat{\Phib}(\xb)$, which is linear in the lifted state, as a special case for $L_w=0$ and $K=LP^{-1}$.
However, our proposed nonlinear controller parametrization~\eqref{eq:controller-parametrization-nonlinear} achieves a larger RoA for the closed-loop system.

On a technical level, the controller design approach in Theorem~\ref{thm:stability-condition-LFR} follows similar ideas as in~\cite{strasser:schaller:worthmann:berberich:allgower:2025}.
The key difference of Theorem~\ref{thm:stability-condition-LFR} to this existing work is that the latter designs a data-driven controller based on an estimate of the Koopman \emph{generator}, i.e., it involves learning a continuous-time model and designing a controller, which continuously changes the respective control values. 
The key disadvantage of the scheme from~\cite{strasser:schaller:worthmann:berberich:allgower:2025} is that derivative measurements are required, which limits its practical use, especially in the presence of noisy data.
Instead, Theorem~\ref{thm:stability-condition-LFR} relies on a discrete-time controller design for the sampled nonlinear system with sample-and-hold implementation of the control signal~\eqref{eq:dynamics-nonlinear-sampled} (compare~\cite{strasser:berberich:allgower:2023a,strasser:berberich:allgower:2023b}).
The main contribution of the introduced controller design is the tight connection to the SafEDMD-based bilinear surrogate and, hence, it leads to closed-loop guarantees for the true nonlinear system purely based on (non-derivative) data.
\begin{corollary}\label{cor:controller-certificates-continuous-time}
    Let the assumptions of Theorem~\ref{thm:stability-condition-LFR} hold and suppose additionally that the vector field $\fb_c:\bbR^n\times\bbR^m \to \bbR^n$ with $\fb_c(\xb,\ub)=\fb(\xb)+\gb(\xb)\ub$ corresponding to the nonlinear system~\eqref{eq:dynamics-nonlinear} is continuous and locally Lipschitz in its first argument around the origin.\\    
    If there exist a matrix $0\prec P=P^\top\in \bbR^{N\times N}$, matrices $L\in\bbR^{m\times N}$, $L_w\in\bbR^{m\times Nm}$, a matrix $0\prec \Lambda=\Lambda^\top \in \bbR^{m\times m}$, and scalars $\nu>0$, $\tau>0$
    such that~\eqref{eq:stability-condition-LFR} and \eqref{eq:stability-condition-LFR-invariance} hold,
    then there exists an amount of data $d_0\in\bbN$ such that for all $d\geq d_0$ the controller
    \begin{equation}\label{eq:controller-sampled}
    \hspace*{-0.0175\linewidth}
        \mub_\mathrm{s}(\xb(t)) 
        = \mub(\xb(k\Delta t)),
        \;
        t\in[k\Delta t,(k+1)\Delta t)
        ,\,
        k\geq 0
    \end{equation}
    ensures exponential stability of the \emph{continuous-time} nonlinear system~\eqref{eq:dynamics-nonlinear} for all initial conditions $\hat{\xb}\in\cX_\mathrm{RoA} := \{\xb\in\bbR^n\,|\, \hat{\Phib}(\xb)^\top P^{-1} \hat{\Phib}(\xb) \leq 1\}$ with probability $1-\delta$.%
\end{corollary}%
\begin{pf}%
    This result can be proven building on~\cite[Thm.~2.27]{grune:pannek:2017}, which is elaborated in Appendix~\ref{app:proof-controller-continuous-time}.
\end{pf}%
This corollary uses the stability guarantees in Theorem~\ref{thm:stability-condition-LFR}, which are obtained for the sampled system~\eqref{eq:dynamics-nonlinear-sampled}, to infer closed-loop guarantees also for the \emph{true} continuous-time nonlinear system~\eqref{eq:dynamics-nonlinear} using sampled feedback. 
Hence, the obtained closed-loop guarantees based on the introduced Koopman-based bilinear surrogate architecture 
hold for both sampled and continuous-time nonlinear systems.
\begin{remark}
    The proposed SafEDMD-based controller design can also be used to design terminal ingredients for a model predictive controller to further enlarge the RoA~\citep{worthmann:strasser:schaller:berberich:allgower:2024}.
\end{remark}

%% file: sec5-numerical-examples.tex
\section{Numerical experiments}\label{sec:numerics}
To evaluate the proposed SafEDMD scheme, we perform a numerical study and compare our results to state-of-the-art Koopman-based control. 
All our simulations are conducted on an i7~notebook using Yalmip~\citep{lofberg:2004} with the semi-definite programming solver MOSEK~\citep{mosek:2022} in Matlab.\footnote{The code is available via \href{https://github.com/rstraesser/SafEDMD}{\texttt{https://github.com/\allowbreak rstraesser/SafEDMD}}.}

\subsection{EDMDc with linear quadratic regulation}\label{sec:numerics:EDMDc}
First, we recap the state-of-the-art Koopman learning EDMDc method for controlled systems with an LQR controller used later for comparison.
As defined in~\cite{brunton:brunton:proctor:kutz:2016,korda:mezic:2018a}, EDMDc relies on the lifted Koopman model 
\begin{equation}\label{eq:surrogate-LQR}
    \hat{\Phib}_+ \approx \check{A} \hat{\Phib} + \check{B} \ub 
\end{equation}
with $\check{A}$ and $\check{B}$ obtained via the linear regression
\begin{multline*}
    \min_{\check{A},\check{B}} \Bigg\|
        \begin{bmatrix}
            \hat{\Phib}(X_+^\zerob) & \hat{\Phib}(X_+^{\eb_1}) & \cdots & \hat{\Phib}(X_+^{\eb_m})
        \end{bmatrix}
        \\
        - \begin{bmatrix} 
            \check{A} & \check{B}
        \end{bmatrix}
        \begin{bmatrix}
            \hat{\Phib}(X^\zerob) & \hat{\Phib}(X^{\eb_1}) & \cdots & \hat{\Phib}(X^{\eb_m}) \\
            0_{m\times d^0} & \mathbf{1}_{d^{\eb_1}}^\top\kron \eb_1 & \cdots & \mathbf{1}_{d^{\eb_m}}^\top\kron \eb_m
        \end{bmatrix}
    \Bigg\|,
\end{multline*}
where $\mathbf{1}_{d}\in\bbR^{d}$ denotes a vector containing ones. 
Based on that, a commonly employed control strategy is the discrete-time LQR controller $\mub_\mathrm{LQR}=\check{K}\hat{\Phib}(\xb)$ for the lifted surrogate model, where 
\begin{equation}\label{eq:LQR-gain}
    \check{K} = -(R+\check{B}^\top \check{P} \check{B})^{-1}\check{B}^\top \check{P}\check{A}
\end{equation}
with $\check{P}$ denoting the solution to the discrete-time algebraic Riccati equation 
\begin{equation*}
    \check{P} = \check{A}^\top \check{P} \check{A} - \check{A}^\top \check{P} \check{B}(R + \check{B}^\top \check{P} \check{B})^{-1}\check{B}^\top \check{P} \check{A} + Q.
\end{equation*}
Here, $Q$ and $R$ need to be positive definite, and we choose $Q=R=I$ for simplicity.
    
We emphasize that the obtained model~\eqref{eq:surrogate-LQR} has no guarantees on its learning quality or error.
As a consequence, the controller $\mub_\mathrm{LQR}$ is not guaranteed a priori to stabilize the origin of the underlying nonlinear system.

\subsection{Nonlinear inverted pendulum}
The inverted pendulum is a classical example for dynamical systems serving as a benchmark in nonlinear data-driven control,~cf.~\cite{bertalan:dietrich:mezic:kevrekidis:2019,chang:roohi:gao:2019,strasser:berberich:allgower:2021,verhoek:abbas:toth:2023,tiwari:nehma:lusch:2023,martin:schon:allgower:2023b}).
The dynamical system 
\begin{equation*}
    \ddot{\theta}(t) = \frac{g}{l}\sin(\theta(t)) - \frac{b}{ml^2}\dot{\theta}(t) + \frac{1}{ml^2}u(t)
\end{equation*}
with mass $m$, length $l$, rotational friction coefficient $b$, and gravitational constant $g=\SI{9.81}{m\per s^2}$
has two state variables $\xb=(\theta,\dot{\theta})$, i.e., the angular position $\theta$ and the angular velocity $\dot{\theta}$, and one control input $u$. Our experiments are performed with $m=1$, $l=1$, and $b=0.01$. 
Further, we define the sets $\bbX=[-2,10]^2$ and $\bbU=[-10,10]$ whereof we uniformly sample $d=\SI{6000}{}$ data points with sampling period $\Delta t=\SI{0.01}{s}$.
More precisely, we consider imperfect measurements to illustrate the practical usability of our approach, i.e., we collect data $\{\xb_j, \tilde{\yb}_j\}_{j=1}^d$ with $\tilde{\yb}_j = \yb_j + \xib_j$, where $\xib_j$ is uniformly sampled from $[-\bar{\xi}, \bar{\xi}]^2$ with $\bar{\xi} = 0.01$.
We use the observables $\hat{\Phib}(\xb) = \begin{bmatrix} \theta & \dot{\theta} & \sin(\theta)\end{bmatrix}^\top$, where the sine function is inspired by prior knowledge on the underlying nonlinear system.
We use the proposed architecture SafEDMD in Algorithm~\ref{alg:learning-architecture} to obtain a data-driven surrogate model~\eqref{eq:dynamics-operator-surrogate-approx} by linear regression.
Moreover, we employ the error bound~\eqref{eq:proportional-bound-error} with $\delta=\SI{0.05}{}$ and $c_x=c_u=\SI{3e-4}{}$, compare Corollary~\ref{cor:proportional-bound-error}.

Based on the bilinear surrogate and its proportional error bounds, we apply the robust controller design in Section~\ref{sec:controller-design} guaranteeing closed-loop exponential stability of the nonlinear system. 
To apply the proposed scheme, we predefine the outer bound $\mathbf{\Delta}_{\Phib}$ on the RoA following the approach in~\cite[Procedure~8]{strasser:schaller:worthmann:berberich:allgower:2025}.
In particular, we first solve~\eqref{eq:stability-condition-LFR} for $\tQ_z=-I$, $\tsb_z=\zerob$, and arbitrary $\tr_z>0$ without considering the second linear matrix inequality~\eqref{eq:stability-condition-LFR-invariance}. 
This leads to a matrix~$P$, which relates the measured data and the chosen observables to maximize the RoA.
Then, we incorporate this information in the actual controller design by defining $\mathbf{\Delta}_{\Phib}$ via $Q_z=-\frac{P^{-1}}{\|P^{-1}\|_2}$, $\sb_z=\zerob$, and $r_z=2.5$.
Designing the nonlinear state-feedback controller~\eqref{eq:controller-parametrization-nonlinear} via Theorem~\ref{thm:stability-condition-LFR} leads to the closed-loop RoA depicted in Fig.~\ref{fig:exmp-inverse-pendulum-sin-comparison}.
The closed-loop trajectories of the nonlinear system highlight that the system is indeed stabilized by our controller inside of the obtained region.
We note that both Algorithm~\ref{alg:learning-architecture} as well as the controller design following Theorem~\ref{thm:stability-condition-LFR} are solved in less than one second.

Next, we compare the bilinear surrogate model using SafEDMD to the state-of-the-art Koopman learning in the literature.
While the literature on learning controllers for nonlinear systems is vast, we emphasize that other results are often achieved via completely different assumptions and provide different theoretical guarantees (if any).
For example,~\cite{chang:roohi:gao:2019} requires initial availability of a local linear quadratic regulator (LQR) as well as measurements of the vector field, i.e., state-derivative data.
Similarly,~\cite{zhang:zhu:lin:2022} requires state-derivative measurements. 
On the contrary, our proposed control works for unknown continuous-time nonlinear systems using only discrete samples of the state, requiring neither derivative data nor a pre-stabilizing controller.
In the following, we compare the proposed approach (using a bilinear surrogate model) to an LQR based on a linear surrogate model, as considered, e.g., in~\cite{brunton:brunton:proctor:kutz:2016,korda:mezic:2018a}, see the recap in Section~\ref{sec:numerics:EDMDc} for details.
Based on this linear model, we design a linear LQR control law $\mu_\mathrm{LQR}=\hat{K}\hat{\Phib}(\xb)$ with matrix $\hat{K}\in\bbR^{1\times N}$ defined in~\eqref{eq:LQR-gain}.
We emphasize that the model obtained by EDMDc does not admit any error bounds, compare~\cite[Sec.~5.3]{iacob:toth:schoukens:2024}.
As a consequence, we observe in Fig.~\ref{fig:exmp-inverse-pendulum-sin-comparison} that $\mu_\mathrm{LQR}$ is not able to stabilize the origin of the underlying nonlinear system.
Moreover, the obtained closed-loop behavior of the nonlinear system is highly sensitive to tuning parameters of the LQR.
On the other hand, the proposed controller design architecture, SafEDMD with a bilinear lifted model and guaranteed learning error bounds, leads to a stabilizing controller for the \emph{unknown} nonlinear system~\eqref{eq:dynamics-nonlinear} using the same data samples.
\begin{figure}[t]
    \centering
    \captionsetup[subfloat]{position=bottom, captionskip=-3pt, nearskip=1pt, farskip=1pt}
    \subfloat[Inverted pendulum.]{\label{fig:exmp-inverse-pendulum-sin-comparison}
        \begin{tikzpicture}[%
            /pgfplots/every axis x label/.style={at={(0.5,0)},yshift=-20pt},%
            /pgfplots/every axis y label/.style={at={(0,0.5)},xshift=-25pt,rotate=0},%
          ]%
            \begin{axis}[
                axis equal,
                legend pos= south east,
                xlabel=$\theta$,
                xmin=-8,
                xmax=8,
                xtick distance=5,
                minor x tick num=4,
                ylabel=$\dot{\theta}$,
                ymin=-9,
                ymax=9,
                ytick distance=5,
                minor y tick num=4,
                grid=both,
                width = 0.85\columnwidth,
                minor grid style={gray!20},
                unbounded coords = jump,
                restrict x to domain =-20:20,
                restrict y to domain =-20:20,
            ]
                \addplot[black,fill=black!40,thick,smooth] table [x index=0,y index=1] {data/exmp-inverse-pendulum-sin-noisy.dat};
                \addplot[black,fill=black!10,thick,smooth] table [x index=4,y index=5] {data/exmp-inverse-pendulum-sin-noisy.dat};
                \addplot[black,thick,smooth] table [x index=14,y index=15] {data/exmp-inverse-pendulum-sin-noisy.dat};
                \addplot[black,thick,smooth,dashdotted] table [x index=16,y index=17] {data/exmp-inverse-pendulum-sin-noisy.dat};
                \addplot[black,thick,smooth] table [x index=20,y index=21] {data/exmp-inverse-pendulum-sin-noisy.dat};
                \addplot[black,thick,smooth,dashdotted] table [x index=22,y index=23] {data/exmp-inverse-pendulum-sin-noisy.dat};
                \addplot[black,thick,smooth] table [x index=26,y index=27] {data/exmp-inverse-pendulum-sin-noisy.dat};
                \addplot[black,thick,smooth,dashdotted] table [x index=28,y index=29] {data/exmp-inverse-pendulum-sin-noisy.dat};
                \addplot[black,thick,smooth] table [x index=32,y index=33] {data/exmp-inverse-pendulum-sin-noisy.dat};
                \addplot[black,thick,smooth,dashdotted] table [x index=34,y index=35] {data/exmp-inverse-pendulum-sin-noisy.dat};
                \addplot[black,thick,smooth] table [x index=38,y index=39] {data/exmp-inverse-pendulum-sin-noisy.dat};
                \addplot[black,thick,smooth,dashdotted] table [x index=40,y index=41] {data/exmp-inverse-pendulum-sin-noisy.dat};
                \addplot[black,thick,smooth] table [x index=44,y index=45] {data/exmp-inverse-pendulum-sin-noisy.dat};
                \addplot[black,thick,smooth,dashdotted] table [x index=46,y index=47] {data/exmp-inverse-pendulum-sin-noisy.dat};
            \end{axis}
        \end{tikzpicture}
    }
    \\
    \subfloat[Koopman invariant observables.]{\label{fig:exmp-cooked-up-eigenfunctions}
        \hspace*{0.03\linewidth}
        \begin{tikzpicture}[%
            /pgfplots/every axis x label/.style={at={(0.5,0)},yshift=-20pt},%
            /pgfplots/every axis y label/.style={at={(0,0.5)},xshift=-25pt,rotate=90},%
          ]%
            \begin{axis}[
                axis equal,
                legend pos= south east,
                xlabel=$x_1$,
                xmin=-15,
                xmax=15,
                xtick distance=10,
                minor x tick num=4,
                ylabel=$x_2$,
                ymin=-18,
                ymax=22,
                ytick distance=10,
                minor y tick num=4,
                grid=both,
                width = 0.85\columnwidth,
                minor grid style={gray!20},
                unbounded coords = jump,
                restrict x to domain =-25:25,
                restrict y to domain =-40:40,
            ]
                \addplot[black,fill=black!10,thick,smooth] table [x index=4,y index=5] {data/exmp-cooked-up-eigenfunctions.dat};
                \addplot[black,thick,smooth] table [x index=14,y index=15] {data/exmp-cooked-up-eigenfunctions.dat};
                \addplot[black,thick,smooth,dashdotted] table [x index=16,y index=17] {data/exmp-cooked-up-eigenfunctions.dat};
                \addplot[black,thick,smooth] table [x index=20,y index=21] {data/exmp-cooked-up-eigenfunctions.dat};
                \addplot[black,thick,smooth,dashdotted] table [x index=22,y index=23] {data/exmp-cooked-up-eigenfunctions.dat};
                \addplot[black,thick,smooth] table [x index=26,y index=27] {data/exmp-cooked-up-eigenfunctions.dat};
                \addplot[black,thick,smooth,dashdotted] table [x index=28,y index=29] {data/exmp-cooked-up-eigenfunctions.dat};
                \addplot[black,thick,smooth] table [x index=32,y index=33] {data/exmp-cooked-up-eigenfunctions.dat};
                \addplot[black,thick,smooth,dashdotted] table [x index=34,y index=35] {data/exmp-cooked-up-eigenfunctions.dat};
                \addplot[black,thick,smooth] table [x index=38,y index=39] {data/exmp-cooked-up-eigenfunctions.dat};
                \addplot[black,thick,smooth,dashdotted] table [x index=40,y index=41] {data/exmp-cooked-up-eigenfunctions.dat};
                \addplot[black,thick,smooth] table [x index=44,y index=45] {data/exmp-cooked-up-eigenfunctions.dat};
                \addplot[black,thick,smooth,dashdotted] table [x index=46,y index=47] {data/exmp-cooked-up-eigenfunctions.dat};
                \addplot[black,thick,smooth] table [x index=50,y index=51] {data/exmp-cooked-up-eigenfunctions.dat};
                \addplot[black,thick,smooth,dashdotted] table [x index=52,y index=53] {data/exmp-cooked-up-eigenfunctions.dat};
                \addplot[black,thick,smooth] table [x index=56,y index=57] {data/exmp-cooked-up-eigenfunctions.dat};
                \addplot[black,thick,smooth,dashdotted] table [x index=58,y index=59] {data/exmp-cooked-up-eigenfunctions.dat};
                \addplot[black,thick,smooth] table [x index=62,y index=63] {data/exmp-cooked-up-eigenfunctions.dat};
                \addplot[black,thick,smooth,dashdotted] table [x index=64,y index=65] {data/exmp-cooked-up-eigenfunctions.dat};
                \addplot[black,thick,smooth] table [x index=68,y index=69] {data/exmp-cooked-up-eigenfunctions.dat};
                \addplot[black,thick,smooth,dashdotted] table [x index=70,y index=71] {data/exmp-cooked-up-eigenfunctions.dat};
            \end{axis}
        \end{tikzpicture}
    }
    \setbox1=\hbox{\begin{tikzpicture}
        \draw[black,fill=black!40,thick](0,0) circle (0.08);
    \end{tikzpicture}}
    \setbox2=\hbox{\begin{tikzpicture}
        \draw[black,fill=black!10,thick](0,0) circle (0.08);
    \end{tikzpicture}}
    \setbox3=\hbox{\begin{tikzpicture}[baseline]
        \draw[black,thick,smooth] (0,.6ex)--++(1.35em,0);
    \end{tikzpicture}}
    \setbox4=\hbox{\begin{tikzpicture}[baseline]
        \draw[black,thick,smooth,dashdotted] (0,.6ex)--++(1.35em,0);
    \end{tikzpicture}}
    \vspace*{-0.75\baselineskip}
    \caption{Region containing all $\xb$ with $\hat{\Phib}(\xb)\in\mathbf{\Delta}_{\Phib}$ (\usebox1), RoA $\cX_\mathrm{RoA}$ (\usebox2), and closed-loop trajectories for the controllers $\mu$ (\usebox3) and $\mu_\mathrm{LQR}$ (\usebox4).}
    \label{fig:exmp}
\end{figure}
\begin{remark}
    Corollary~\ref{cor:proportional-bound-error} guarantees the error bound~\eqref{eq:proportional-bound-error} for a user-selected probabilistic tolerance $\delta$, sampling time $\Delta t$, and data length $d$ (assuming the projection error bound~\eqref{eq:proportional-bound-projection}). 
    Since the error bounds are valid for a general class of nonlinear systems, the resulting constants $c_x$, $c_u$ may be conservative, and, as demonstrated, even smaller constants may accurately capture the approximation error of the data-driven surrogate of the true system. 
    Hence, we use user-selected $c_x$, $c_u$ in the examples and propose an iterative procedure between controller design and data collection. 
    In particular, a robust controller for larger constants yields closed-loop guarantees for a broader class of nonlinear systems.
    Possible ideas to reduce the conservatism and sharpen the data requirements can be found in~\cite{philipp:schaller:boshoff:peitz:nuske:worthmann:2024}.
\end{remark}

To illustrate how different error bounds influence the resulting RoA, Fig.~\ref{fig:controller-feasiblity} depicts for a fixed $\mathbf{\Delta}_{\Phib}$ the size of the established RoA in terms of $\mathrm{tr}(P^{-1})$ depending on the values of $c_x = c_u$ in the proportional error bound~\eqref{eq:proportional-bound-error}.
Here, a smaller value of $\mathrm{tr}(P^{-1})$ corresponds to a larger RoA. 
As expected, the established RoA using a robust controller shrinks as the size of the error bound increases.
\begin{figure}[t]
    \centering
    \input{figs/feasibility-comparison}
    \vspace*{-0.5\baselineskip}
    \caption{Relationship between $\mathrm{tr}(P^{-1})$ (as a measure of the size of $\cX_\mathrm{RoA}$) and different proportional error bounds with $c_x=c_u$, averaged over 20 runs.}
    \label{fig:controller-feasiblity}
\end{figure}

\subsection{Nonlinear system with Koopman invariant observables}
Next, we consider a common nonlinear benchmark system which admits an invariant dictionary~\citep{brunton:brunton:proctor:kutz:2016}.
The system has a two-dimensional state and one input, and the dynamics are given by
\begin{align*}
    \dot{x}_1(t) 
    &= \rho x_1(t), 
    \\
    \dot{x}_2(t) 
    &= \lambda (x_2(t) - x_1(t)^2) + u(t)
\end{align*}
with $\rho,\lambda\in\bbR$.
Our simulations are conducted for $\rho=-2$ and $\lambda=1$.
This example permits the definition of an invariant dictionary~\citep{brunton:brunton:proctor:kutz:2016} via the observable function
$
    \hat{\Phib}(\xb) = \begin{bmatrix}
        x_1 & x_2 & x_2-\frac{\lambda}{\lambda-2\rho}x_1^2
    \end{bmatrix}^\top
$
leading to \emph{exact} finite-dimensional lifted continuous-time dynamics
\begin{equation}\label{eq:exmp-cooked-invariant}
    \ddt{}\hat{\Phib}(\xb(t)) = \begin{bmatrix}
        \rho & 0 & 0 \\
        0 & 2\rho & \lambda-2\rho \\
        0 & 0 & \lambda
    \end{bmatrix}
    \hat{\Phib}(\xb(t)) 
    + \begin{bmatrix}
        0 \\ 1 \\ 1
    \end{bmatrix}
    u(t).
\end{equation}
Here, even though $\Phib$ is Koopman-invariant and~\eqref{eq:exmp-cooked-invariant} is exact, the formulation is exposed to a major drawback: The lifted dynamics are time-continuous, i.e., information on the state derivative is necessary in order to learn the exact continuous-time dynamics from data. 
In particular, the derivative data either needs to be measured directly or estimated based on sample data.
This, however, introduces errors (in particular in the presence of noise) and, thus, prevents safe and reliable learning of the underlying true system. 
Contrary, the proposed SafEDMD scheme provides a reliable surrogate model purely based on sample data of state $x$ and, particularly, without the need for derivative data.

For the data generation, we sample $d=100$ data points uniformly from the sets $\bbX=[-1,1]^2$ and $\bbU=[-1,1]$ with sampling rate $\Delta t=\SI{0.01}{s}$.
Then, we learn a data-driven bilinear surrogate model via SafEDMD based on the generated data set and, particularly, without the need for derivative data.
For a probabilistic tolerance $\delta=\SI{0.05}{}$ and constants $c_x=c_u=\SI{5e-3}{}$ for the error bound~\eqref{eq:proportional-bound-error}, we apply the controller design in Theorem~\ref{thm:stability-condition-LFR} with $\mathbf{\Delta}_{\Phib}$ defined by $Q_z=-I$, $\sb_z=\zerob$, and $r_z=\SI{500}{}$.
The obtained controller is guaranteed to ensure the RoA $\cX_\mathrm{RoA}$ shown in Fig.~\ref{fig:exmp-cooked-up-eigenfunctions}. 
Here, $\cX_\mathrm{RoA}$ contains all $\xb$ with $\hat{\Phib}(\xb)\in\mathbf{\Delta}_{\Phib}$ and, thus, adds no additional conservatism to the robust controller design.
As before, both the learning and the controller design are completed in less than one second.
In summary, also for this example, the controlled system under the proposed controller $\mu$ exponentially converges to the origin.
\begin{remark}
    Although the considered examples are low-dimensional, we already see significant advantages over state-of-the-art EDMDc.
    In particular, EDMDc fails to stabilize the underlying system due to the lack of closed-loop guarantees.
    In future work, we aim to apply the developed results to larger systems, where it is crucial to exploit structural properties of the LMI feasibility problem (see the discussion after Theorem~\ref{thm:stability-condition-LFR}). 
    Further, the predefined choice of $\mathbf{\Delta}_{\mathbf{\Phi}}$ may lead to a conservative RoA in practice. 
    To address this issue,~\cite{strasser:berberich:allgower:2025,strasser:berberich:schaller:worthmann:allgower:2025} propose a sum-of-squares optimization-based controller design that directly exploits the bilinearity of the surrogate dynamics and, thus, significantly increases the obtained RoA.
\end{remark}

%% file: figs/feasibility-comparison.tex
\begin{tikzpicture}[%
/pgfplots/every axis y label/.style={at={(0,0.5)},xshift=-35pt,rotate=90},%
]%
\begin{axis}[%
width=0.85\columnwidth,
height=0.325\columnwidth,
xmode=log,
xmin=1e-05,
xmax=0.01,
xlabel={$c_x=c_u$},
xminorticks=true,
ymin=0.45,
ymax=0.55,
ytick distance=0.05,
ylabel={$\mathrm{tr}(P^{-1})$},
yminorticks=true,
xmajorgrids,
xminorgrids,
ymajorgrids,
yminorgrids
]
\addplot [color=black, thick, forget plot]
  table[row sep=crcr]{%
0.01	0.533773457826612\\
0.00899999999999999	0.519412167473768\\
0.008	0.512125025576567\\
0.007	0.499035617517204\\
0.006	0.495221205923653\\
0.005	0.489853075686736\\
0.004	0.487117543989487\\
0.003	0.483173184773266\\
0.002	0.481491927238988\\
0.001	0.480568883480571\\
0.000899999999999999	0.483808272855148\\
0.0008	0.48238578494267\\
0.0007	0.480261340965728\\
0.0006	0.483236086936221\\
0.0005	0.480048736126097\\
0.0004	0.481387399156169\\
0.0003	0.48061006568033\\
0.0002	0.479800351979364\\
0.0001	0.476313101072743\\
1e-05	0.47468763147912\\
};
\end{axis}
\end{tikzpicture}%

%% file: sec6-conclusion.tex
\section{Conclusion and outlook}\label{sec:conclusion}
In conclusion, this paper introduced SafEDMD, a Koopman-based controller design framework tailored to data-driven control of nonlinear dynamical systems.
Leveraging the analytical foundation of the Koopman operator, we showed that SafEDMD enables a rigorous derivation of error bounds crucial for robust controller design. 
Unlike existing methods, the proposed architecture handles the Koopman operator directly via sampled data, eliminating the need for hard-to-obtain time-derivative data. 
SafEDMD offers a promising direction for the control of unknown nonlinear systems with closed-loop guarantees, demonstrating its superiority in scenarios where existing methods fail. 
This research not only tailors the capabilities of bilinear EDMD for control tasks but also establishes a new paradigm in the field of learning and control of data-driven surrogate models with rigorous stability guarantees, addressing a critical aspect in safety-critical applications.

Interesting future work includes the integration of deep NNs as the observable function~\citep{takeishi:kawahara:yairi:2017,yeung:kundu:hodas:2019,shi:meng:2022} to exploit the superior robustness and scalability of EDMD-based approaches compared to state-of-the-art reinforcement learning methods (compare~\cite{han:euler-rolle:kratzschmann:2021}).
Further, extending a posteriori bounds based on residual dynamic mode decomposition~\citep{colbrook:townsend:2024} for autonomous systems to controlled systems may provide insights into the tightness of the derived a priori bounds.
Last, kernel techniques combined with spectral methods for estimating the Koopman generator~\citep{rosenfeld:kamalapurkar:2024} may be beneficial for a non-conservative controller design due to the absence of projection and time-discretization errors. 
These alternative operator learning methods mentioned above can replace the (non-proportional) bound on the \emph{autonomous} Koopman operator to derive a proportional bound as in Corollary~\ref{cor:proportional-bound-error} via the SafEDMD framework, enabling data-driven controller design with closed-loop guarantees.

%% file: secAppendix-proof-proportional-bound-DD-surrogate.tex
\section{Proof of Theorem~\ref{thm:proportional-bound-data}}\label{app:proof-proportional-bound-DD-surrogate}

In this section, we prove Theorem~\ref{thm:proportional-bound-data}, which builds the foundation of SafEDMD: A novel Koopman-based framework for data-driven control of nonlinear dynamical systems.
To this end, we first introduce a finite-dimensional bilinear representation for the projected Koopman operator in Lemma~\ref{lm:proportional-bound-bilinear-representation} before defining a data-driven surrogate model for the bilinear representation in Lemma~\ref{lm:proportional-bound-DD-surrogate}. 
Then, Theorem~\ref{thm:proportional-bound-data} results as a direct consequence of the preliminary lemmas.

\subsection{Bilinear representation for the Koopman operator}\label{sec:bilinear-representation-Koopman}
The proposed architecture relies on a bilinear representation of the projected Koopman operator $P_\bbV\cK_{\Delta t}^\ub|_\bbV$.
For the setting in this paper and for a fixed control value $\ub\in \bbU$, $(\cK_t^\ub)_{t\geq 0}$ is a strongly-continuous semigroup of bounded linear operators, i.e., we can define its infinitesimal generator $\cL^\ub$ via  
\begin{equation*}
    \cL^\ub \varphi \coloneqq \lim_{t\searrow 0} \frac{\cK_t^\ub\varphi - \varphi}{t}
    \quad 
    \forall\, \varphi \in D(\cL^\ub),
\end{equation*}
where the domain $D(\cL^\ub)$ consists of all $L^2$-functions for which the above limit exists. 
By definition of this generator, the propagated observable satisfies $\dot{\varphi}(\xb)=\cL^\ub \varphi(\xb)$.
A key structural property of the Koopman generator, also called Liouville operator, is that control affinity is preserved~\citep{williams:hemati:dawson:kevrekidis:rowley:2016,surana:2016}, i.e., for $\ub\in\bbR^m$ we have
\begin{equation}\label{eq:Koopman-generator}
    \cL^\ub = \cL^\zerob + \sum_{i=1}^m u_i (\cL^{\eb_i}-\cL^\zerob),
\end{equation}
where $\cL^\zerob$ and $\cL^{\eb_i}$, $i\in[m]$, are the generators of the Koopman semigroups corresponding to the constant control functions $\ub\equiv \zerob$ and $\ub\equiv \eb_i$ with the unit vector $\eb_i$, $i\in[m]$, respectively.
Hence, the Koopman generator acting on the vector-valued observable function~$\Phib$ in~\eqref{eq:lifting-function} has a specific structure due to the included constant observable $\phi_0\equiv 1$. 
In particular, we use~\eqref{eq:Koopman-generator} to obtain
\begin{gather}\label{eq:structure-Koopman-generator}
    \cL^\ub = \begin{bmatrix}
        0 & \zerob^\top \\ \cL^{\ub}_{21} & \cL^{\ub}_{22} 
    \end{bmatrix}
    ,\;
    \cL^\zerob = \begin{bmatrix}
        0 & \zerob^\top \\ \zerob & \cL^\zerob_{22}
    \end{bmatrix}
    ,\;
    \cL^{\eb_i} = \begin{bmatrix}
        0 & \zerob^\top \\ \cL^{\eb_i}_{21} & \cL^{\eb_i}_{22}
    \end{bmatrix},
\end{gather}
$i\in[m]$.
The structure of the 
matrices in~\eqref{eq:structure-Koopman-generator} corresponds to the structure of the Koopman operator~\eqref{eq:structure-Koopman-operator} in the proposed learning architecture. 
We stress that in this work, the Koopman generator is only used in the proofs.
In particular, our learning scheme directly approximates the Koopman operator by means of trajectory samples. 
Existing Koopman-based controller design methods with guarantees are mostly based on (typically extremely-noisy) derivative data to approximate the Koopman generator in view of $\mathcal{L}^\ub \Phib(\xb) = \langle \nabla \Phib(\xb), \dot{\xb}(t) \rangle = \langle \nabla \Phib(\xb), \fb(\xb(t)) + \sum_{i=1}^m \gb_i(\xb(t)) u_i(t) \rangle$. 
In conclusion, this makes the learning scheme of our controller design framework data-efficient and robust in comparison to competing methods, for which similar error estimates exist~\citep{bold:grune:schaller:worthmann:2025,strasser:schaller:worthmann:berberich:allgower:2025} and whose derivations have inspired the following lemma and its proof.
We note that a promising alternative is the spectral approach of~\cite{rosenfeld:kamalapurkar:2024} to approximate the Koopman generator from trajectory samples. 
While this approach focuses on the \emph{prediction} of a closed-loop dynamics, using it in controller \emph{design} is an interesting issue for future research.

In the following, we leverage the control affine structure~\eqref{eq:Koopman-generator} of the generator $\cL^\ub$ to study the Koopman operator given by $P_\bbV \cK_{\Delta t}^\ub|_\bbV = \exp(\Delta t P_\bbV\cL^\ub|_\bbV)$.
To this end, we recall~\eqref{eq:Koopman-bilinear-approximate} and consider the projected lifted dynamics
\begin{multline*}
    {(P_\bbV\cK_{\Delta t}^\ub|_\bbV)} \Phib(\xb)
    = {(P_\bbV\cK_{\Delta t}^\zerob|_\bbV)} \Phib(\xb) 
    \\
    + \sum_{i=1}^m u_{i}({P_\bbV\cK_{\Delta t}^{\eb_i}|_\bbV} - {P_\bbV\cK_{\Delta t}^\zerob|_\bbV})\Phib(\xb) 
    + \xib(\xb, \ub)
\end{multline*}
of the projected Koopman operator $P_\bbV\cK_{\Delta t}^\ub|_\bbV$ on the dictionary $\bbV$ with approximation error $\xib(\xb, \ub)\coloneqq h(\Delta t)\Phib(\xb)$ for
\begin{multline}\label{eq:bilinear-representation-h}
    h(\Delta t) 
    \coloneqq 
    {P_\bbV\cK_{\Delta t}^\ub|_\bbV} 
    \\
    - \left[{P_\bbV\cK_{\Delta t}^\zerob|_\bbV} + \sum_{i=1}^m u_i({P_\bbV\cK_{\Delta t}^{\eb_i}|_\bbV} - {P_\bbV\cK_{\Delta t}^\zerob|_\bbV)}\right].
\end{multline}
The following lemma formulates a proportional bound on the approximation error $\xib$ of the bilinear representation of the Koopman operator.
\begin{lemma}\label{lm:proportional-bound-bilinear-representation}
    The approximation error $\xib$ due to the bilinear representation of the projected Koopman operator $P_\bbV\cK_{\Delta t}^\ub|_\bbV$ satisfies the proportional bound
    \begin{equation}\label{eq:proportional-bound-bilinear-representation}
        \|\xib(\xb,\ub)\| \leq \Delta t^2 (c_1\|\Phib(\xb)-\Phib(\zerob)\| + \sqrt{m}c_2 \|\ub\|),
    \end{equation}
    where $c_1,c_2\in\bbR_+$ are defined in~\eqref{eq:estimate-c1} and~\eqref{eq:estimate-c2}.
\end{lemma}
\begin{pf}
    Recall the definition of $\xib(\xb,\ub)=h(\Delta t)\Phib(\xb)$ with $h(\Delta t)$ defined in~\eqref{eq:bilinear-representation-h}.
    Then, we bound the norm of $\xib(\xb,\ub)$ by
    \begin{align}\label{eq:bound-xi-individual-terms}
        \|\xib(\xb,\ub)\|
        &= \|h(\Delta t)(\Phib(\xb) - \Phib(\zerob) + \Phib(\zerob))\| 
        \nonumber \\
        &\leq \|h(\Delta t)(\Phib(\xb)-\Phib(\zerob))\| + \|h(\Delta t)\Phib(\zerob)\|
        \nonumber \\
        &\leq \|h(\Delta t)\|\|\Phib(\xb)-\Phib(\zerob)\| + \|h(\Delta t)\Phib(\zerob)\|.
    \end{align}
    To estimate the first term in~\eqref{eq:bound-xi-individual-terms}, we substitute $P_\bbV\cK_{\Delta t}^\ub|_\bbV = e^{\Delta tP_\bbV\cL^\ub|_\bbV}$, $P_\bbV\cK_{\Delta t}^\zerob|_\bbV = e^{\Delta t{P_\bbV\cL^\zerob|_\bbV}}$, and ${P_\bbV\cK_{\Delta t}^{\eb_i}|_\bbV} = e^{\Delta t{P_\bbV\cL^{\eb_i}|_\bbV}}$, $i\in[m]$, to obtain 
    \begin{multline}\label{eq:Taylor-h(Delta t)}
    	\hspace*{-0.025\linewidth}
        h(\Delta t) 
        = e^{
            \Delta t\left(
                {P_\bbV\cL^\zerob|_\bbV} + \sum_{i=1}^m u_i({P_\bbV\cL^{\eb_i}|_\bbV}-{P_\bbV\cL^\zerob|_\bbV})
            \right)
        } 
        \\\hspace*{-0.025\linewidth}
        - \Big(
            e^{\Delta t{P_\bbV\cL^\zerob|_\bbV}} + \sum_{i=1}^m u_i\big(e^{\Delta t{P_\bbV\cL^{\eb_i}|_\bbV}} - e^{\Delta t{P_\bbV\cL^\zerob|_\bbV}}\big)
        \Big).\!\!\!\!
    \end{multline}
    Then, the Taylor series expansion for $h(\Delta t)$ at $\Delta t=0$ yields, in view of $h(0) = 0$ and $\frac{\partial h}{\partial \Delta t}(0) = 0$,
    \begin{equation*}
        h(\Delta t) 
        = h(0) + \Delta t \frac{\partial h}{\partial \Delta t}(0) + \frac{\Delta t^2}{2} \frac{\partial^2h}{\partial\Delta t^2}(\tau)
        = \frac{\Delta t^2}{2} \frac{\partial^2h}{\partial\Delta t^2}(\tau)
    \end{equation*}
    for some $\tau\in[0,\Delta t]$, which is an exact representation of $h(\Delta t)$ since we include the Lagrange remainder evaluated at $\tau$. 
    Then, $
        h(\Delta t) 
        = \tfrac{\Delta t^2}{2}(h_1(\tau)h_2(\tau)^2 - h_3(\tau))
    $, where
    \begin{align*}
        h_1(\tau) 
        &= e^{
            \tau\left(
                {P_\bbV\cL^\zerob|_\bbV} + \sum_{i=1}^m u_i({P_\bbV\cL^{\eb_i}|_\bbV}-{P_\bbV\cL^\zerob|_\bbV})
            \right)
        },
        \\
        h_2(\tau) 
        &= {P_\bbV\cL^\zerob|_\bbV} 
        + \sum_{i=1}^m u_i({P_\bbV\cL^{\eb_i}|_\bbV}-{P_\bbV\cL^\zerob|_\bbV}),
        \\
        h_3(\tau) 
        &= 
        e^{\tau{P_\bbV\cL^\zerob|_\bbV}}({P_\bbV\cL^\zerob|_\bbV})^2
        + \sum_{i=1}^m u_i\big(
            e^{\tau{P_\bbV\cL^{\eb_i}|_\bbV}}({P_\bbV\cL^{\eb_i}|_\bbV})^2 
        \\&\qquad\qquad\qquad\qquad\qquad
            - e^{\tau{P_\bbV\cL^\zerob|_\bbV}}({P_\bbV\cL^\zerob|_\bbV})^2
        \big).
    \end{align*}
    Hence, we can derive the following estimate using the triangle inequality 
    \begin{align}\label{eq:estimate-c1}
   	\hspace*{-0.03\linewidth}
        \frac {2 \| h(\Delta t) \|} {\Delta t^2} 
        &\leq 
        \left\|
            h_1(\tau)
        \right\|
        \left\|
            h_2(\tau)
        \right\|^2
        + \left\|
            h_3(\tau)
        \right\|
        \nonumber\\
        &\leq
        e^{
            \Delta t \left(
                \hat{u}
                \| {P_\bbV\cL^\zerob|_\bbV} \|
                + \max_{\ub\in\bbU} \|u\|_1 \| \cL^\star \|
            \right)
        }
        \nonumber\\
        &\qquad \cdot
        \Big(
            \hat{u}
            \|{P_\bbV\cL^\zerob|_\bbV}\| + \max_{\ub\in\bbU} \|u\|_1 \| \cL^\star \|
        \Big)^2 
        \nonumber\\&
        + \hat{u}
        e^{\Delta t \|{P_\bbV\cL^\zerob|_\bbV}\|}
        \left\|
            {P_\bbV\cL^\zerob|_\bbV}
        \right\|^2
        \nonumber\\&
        + \max_{\ub\in\bbU} \|u\|_1 e^{\Delta t \| \cL^\star \|} \| \cL^\star \|^2
        \eqqcolon 2c_1\!\!\!
    \end{align}
    with $\hat{u} := \max_{\ub \in \mathbb{U}} |1 - \sum_{i=1}^m u_i|$ and $
        \| \cL^\star \| \coloneqq$\linebreak $\max_{i\in[m]} \| {P_\bbV\cL^{\eb_i}|_\bbV} \|
    $.
    The constant $c_1$ is finite for a compact set~$\bbU$ of control actions. Hence, we have $\| h(\Delta t) \| \leq \Delta t^2 c_1 < \infty$.
    The terms $\|{P_\bbV\cL^{\bar{\ub}}|_\bbV}\|$ with $\bar{\ub}\in\{\zerob,\eb_1,...,\eb_m\}$ can be further bounded in terms of the system dynamics.
    To see this, we plug in the definition of the operator norm of ${P_\bbV\cL|_\bbV}$ as a mapping from $\mathbb{V}$ to $\mathbb{V}$, where we endow the finite-dimensional subspace $\mathbb{V}$ of $L^2(\bbX)$ with the $L^2$-norm.
    W.l.o.g. the basis elements are scaled such that $\|\phi_\ell\|_{L^2(\bbX)}=1$, $\ell\in[0:N]$, holds.
    Then, we obtain
    \begin{align}
        \|{P_\bbV\cL^{\zerob}|_\bbV}\| 
        &= \sup_{\substack{\varphib\in \mathbb{V},\\\|\varphib\|_{L^2(\bbX)}=1}} \|\nabla \varphib(\cdot)^\top \fb(\cdot)\|_{L^2(\bbX)} 
        \nonumber\\
        &= \sup_{\ell\in[0:N]} \|\nabla \phi_\ell(\cdot)^\top \fb(\cdot)\|_{L^2(\bbX)} 
        \nonumber\\
        &\leq \sup_{\ell\in[0:N]}\|\nabla \phi_\ell\|_{L^2(\bbX)}\|\fb\|_{L^2(\bbX)}
    \label{eq:generator-bounds-L2}
    \end{align}
    and, analogously, for $i\in[m]$ holds
    \begin{equation}
        \|{P_\bbV\cL^{\eb_i}|_\bbV}\| \leq \sup_{\ell\in[0:N]}\|\nabla \phi_\ell\|_{L^2(\bbX)}\|\fb+\gb_i\|_{L^2(\bbX)}.
    \end{equation}
    For an estimate on the second term in~\eqref{eq:bound-xi-individual-terms}, we again use the Taylor expansion~\eqref{eq:Taylor-h(Delta t)}, the triangle inequality, and the notation and arguments used to derive Inequality~\eqref{eq:generator-bounds-L2} to get the estimate
    \vspace*{-2.5\baselineskip}\small%
    \begin{align*}
        &{2 \| h(\Delta t) \Phib(\zerob) \|}/ {\Delta t^2} \\
        &\leq \tilde{c} \bigg\| 
            \Big( 1 - \sum_{i=1}^m u_i \Big)
            \underbrace{({P_\bbV\cL^\zerob|_\bbV}) \Phib(\zerob)}_{=0} 
            + \sum_{i=1}^m u_i \underbrace{({P_\bbV\cL^{\eb_i}|_\bbV)} \Phib(\zerob)}_{= ({P_\bbV\cL^{\eb_i}|_\bbV})_{21}} 
        \bigg\| 
        \\
        &\qquad
        +\bigg\| 
            \bigg( 
                \Big(
                    1 - \sum_{i=1}^m u_i 
                \Big) 
                e^{\tau{P_\bbV\cL^\zerob|_\bbV}} {P_\bbV\cL^\zerob|_\bbV} 
            \bigg) 
            \underbrace{({P_\bbV\cL^\zerob|_\bbV}) \Phib(\zerob)}_{=0} 
        \bigg\| 
        \\
        &\qquad 
        + \Big\| 
            \sum_{i=1}^m u_i e^{\tau{P_\bbV\cL^{\eb_i}|_\bbV}}({P_\bbV\cL^{\eb_i}|_\bbV})^2 \Phib(\zerob) 
        \Big\| 
        \\
        &= \tilde{c} 
        \left\|
            \begin{bmatrix}
                ({P_\bbV\cL^{\eb_1}|_\bbV})_{21} 
                & \cdots 
                & ({P_\bbV\cL^{\eb_m}|_\bbV})_{21}    
            \end{bmatrix}
        \right\| 
        \|\ub\|
        + 
        \left\|\tilde{\cL}\right\| 
        \|\ub\|,
    \end{align*}
    \normalsize
    where
    \begin{align*}
        \tilde{c} 
        &= 
        e^{
            \Delta t \left(
                \hat{u}
                \| {P_\bbV\cL^\zerob|_\bbV} \|
                + \max_{\ub\in\bbU} \|u\|_1 \| \cL^\star \|
            \right)
        } 
        \\
        &\qquad\qquad\qquad
        \cdot
        \Big(
            \hat{u}\|{P_\bbV\cL^\zerob|_\bbV}\| 
            + \max_{\ub\in\bbU} \|u\|_1 \| \cL^\star \|
        \Big),\\
        \tilde{\cL}
        &= 
        \left[\begin{smallmatrix}
            e^{\tau{P_\bbV\cL^{\eb_1}|_\bbV}} ({P_\bbV\cL^{\eb_1}|_\bbV})^2 \Phib(\zerob) 
            & \cdots 
            & e^{\tau{P_\bbV\cL^{\eb_m}|_\bbV}}({P_\bbV\cL^{\eb_m}|_\bbV})^2 \Phib(\zerob)     
        \end{smallmatrix}\right].
    \end{align*}
    Then, we deduce%
    \vspace*{-2\baselineskip}\small%
    \begin{align}
        &{2 \| h(\Delta t) \Phib(\zerob) \|}/{\Delta t^2} 
        \nonumber\\
        &\leq \tilde{c} 
        \sqrt{
            \sum_{i=1}^m \left\|
                ({P_\bbV\cL^{\eb_i}|_\bbV})_{21}
            \right\|^2
        } 
        \|\ub\| 
        \nonumber\\
        &\quad
        + 
        \sqrt{
            \sum_{i=1}^m \left(
                e^{\Delta t\|{P_\bbV\cL^{\eb_i}|_\bbV}\|} \|({P_\bbV\cL^{\eb_i}|_\bbV})_{22}\| \|({P_\bbV\cL^{\eb_i}|_\bbV})_{21}\|
            \right)^2
        } 
        \|\ub\| 
        \nonumber\\
        &\leq \sqrt{m} \big(\tilde{c} \| \cL^\star \| 
        + e^{\Delta t \| \cL^* \|} \| \cL^\star \|^2 \big) \|\ub \|
        \eqqcolon 2\sqrt{m}c_2 \| \ub \|
    \label{eq:estimate-c2}
    \end{align}
    \normalsize
    using $
        ({P_\bbV\cL^{\eb_i}|_\bbV})^2 \Phib(\zerob)=({P_\bbV\cL^{\eb_i}|_\bbV})_{22}({P_\bbV\cL^{\eb_i}|_\bbV})_{21}
    $. 
    Thus, we have $\| h(\Delta t)\Phi(0) \| \leq \Delta t^2 \sqrt{m} c_2 \| \ub \|$.

    Hence, combining the two derived estimates~\eqref{eq:estimate-c1} and~\eqref{eq:estimate-c2}
    yields $\|\xib(\xb,\ub)\| \leq \Delta t^2 (c_1 \| \Phib(\xb)-\Phib(\zerob) \| + \sqrt{m} c_2 \| \ub \|)$, i.e., the desired inequality.
\end{pf}
Here, we emphasize that the constants $c_1,c_2$ in Lemma~\ref{lm:proportional-bound-bilinear-representation} can be computed if, e.g., a bound on the $L^2$-norm of the system dynamics is given.

\subsection{Data-driven surrogate of the bilinear representation of the Koopman operator}\label{sec:data-driven-surrogate-Koopman}
Next, we approximate the bilinear representation of the projected Koopman operator by its respective data-driven estimates.
To this end, we consider the data-driven surrogate model
\begin{align*}
    &({P_\bbV\cK_{\Delta t}^\ub|_\bbV}) \Phib(\xb)
    = \cK_{\Delta t,d}^\ub \Phib(\xb)
    + \xib(\xb, \ub) + \etab(\xb,\ub)
    \\
    &= \cK_{\Delta t,d}^\zerob \Phib(\xb) 
    + \sum_{i=1}^m \ub_i(\cK_{\Delta t,d}^{\eb_i} - \cK_{\Delta t,d}^\zerob)\Phib(\xb) 
    \\&\qquad\qquad\qquad\qquad\qquad\qquad\quad
    + \xib(\xb, \ub) + \etab(\xb,\ub) 
\end{align*}
corresponding to~\eqref{eq:dynamics-operator-surrogate-approx}, where%
\vspace*{-2.5\baselineskip}\small%
\begin{align}\label{eq:eta-definition}
    &\etab(\xb,\ub) 
    \\
    &\quad= ({P_\bbV\cK_{\Delta t}^\zerob|_\bbV}) \Phib(\xb)
    + \sum_{i=1}^m u_{i}({P_\bbV\cK_{\Delta t}^{\eb_i}|_\bbV} - {P_\bbV\cK_{\Delta t}^\zerob|_\bbV})\Phib(\xb)
    \nonumber\\&\quad\qquad 
    - \Big(
        \cK_{\Delta t,d}^\zerob \Phib(\xb) 
        + \sum_{i=1}^m u_{i}(\cK_{\Delta t,d}^{\eb_i} - \cK_{\Delta t,d}^\zerob)\Phib(\xb)
    \Big)
    \nonumber\\
    &\quad= \Big(
        ({P_\bbV\cK_{\Delta t}^\zerob|_\bbV} - \cK_{\Delta t,d}^\zerob)
        + \sum_{i=1}^m u_i ({P_\bbV\cK_{\Delta t}^{\eb_i}|_\bbV} - \cK_{\Delta t,d}^{\eb_i})
    \nonumber\\&\qquad\qquad\qquad
        + \sum_{i=1}^m u_i (\cK_{\Delta t,d}^{\zerob} - {P_\bbV\cK_{\Delta t}^{\zerob}|_\bbV})
    \Big)
    \Phib(\xb)
\end{align}
\normalsize
describes the estimation error due to the learning via sampled data.
Before stating a bound on~$\etab$, we first introduce the following proposition from the literature.
\begin{proposition}[{\citeauthor{nuske:peitz:philipp:schaller:worthmann:2023},~\citeyear{nuske:peitz:philipp:schaller:worthmann:2023},~Thm.~14}]\label{prop:error-bound-operator}
    Suppose that the data samples are i.i.d. Further, let an error bound $c_\eta > 0$ and a probabilistic tolerance $\delta \in (0, 1)$ be given. Then, there is an amount of data $d_0^{\bar{u}} =  \mathcal{O}\left(\nicefrac{1}{\delta c_\eta^2}\right)$ such that for all $d \geq d_0^{\bar{u}}$ we have the error bound
    \begin{equation}\label{eq:error-bound-operator}
        \|P_\bbV\cK_{\Delta t}^{\bar{\ub}}|_\bbV - \cK_{\Delta t,d}^{\bar{\ub}}\| \leq c_\eta
    \end{equation}
    for a fixed control input~$\bar{u}$ with probability $1-\delta$.
\end{proposition}
Now, we show that the sampling error $\etab$ in~\eqref{eq:eta-definition} satisfies the following proportional bound, where we exploit that Proposition~\ref{prop:error-bound-operator} can be applied for any $\bar{\ub}\in\{\zerob,\eb_1,...,\eb_m\}$.
\begin{lemma}\label{lm:proportional-bound-DD-surrogate}
    Suppose that the data samples are i.i.d. 
    Further, let an error bound $c_\eta > 0$ and a probabilistic tolerance $\delta \in (0, 1)$ be given. 
    Then, there is an amount of data $d_0 = \mathcal{O}\left(\nicefrac{1}{\delta c_\eta^2}\right)$ such that for all $d \geq d_0$ the approximation error $\etab$ due to sampling of the Koopman operators $\cK_{\Delta t,d}^{\bar{\ub}}$, $\bar{\ub}\in\{\zerob,\eb_1,...,\eb_m\}$, based on the data satisfies the proportional bound
    \begin{equation}\label{eq:proportional-bound-DD-surrogate}
        \|\etab(\xb,\ub)\| \leq c_3c_\eta \|\Phib(\xb)-\Phib(\zerob)\| + \sqrt{m}c_\eta \|\ub\|
    \end{equation}
    with probability $1-\delta$, where $c_3\in\bbR_+$ is defined in~\eqref{eq:error-estimate-c3-c4}.
\end{lemma}
\begin{pf}
    First, we apply Proposition~\ref{prop:error-bound-operator} for all $\bar{\ub}\in\{\zerob,\eb_1,...,\eb_m\}$ and the given error bound $c_\eta$ and probabilistic tolerance~$\delta$. 
    This results in a necessary amount of data $d_0^{\bar{\ub}} \in \bbN$ for each fixed control input~$\bar{\ub}$ such that the error bound~\eqref{eq:error-bound-operator} holds. 
    We define the maximum of those necessary data lengths as $d_0=\max\{d_0^\zerob,d_0^{\eb_1},...,d_0^{\eb_m}\}$. This ensures that the bound~\eqref{eq:error-bound-operator} holds for all data-driven estimates~$\cK_{\Delta t,d}^{\bar{\ub}}$ of the Koopman operator, where $\bar{\ub} \in \{\zerob,\eb_1,...,\eb_m\}$. 
    Then, we recall the definition of the sampling error $\etab$ in~\eqref{eq:eta-definition} and leverage $\Phib(\xb)=\Phib(\xb)-\Phib(\zerob)+\Phib(\zerob)$ to obtain with $\hat{u} := \max_{\ub \in \mathbb{U}} |1 - \sum_{i=1}^m u_i|$
    \vspace*{-2\baselineskip}\small
    \begin{align*}
        & \|\etab(\xb,\ub)\| 
        \leq \hat{u} \|({P_\bbV\cK_{\Delta t}^\zerob|_\bbV} - \cK_{\Delta t,d}^\zerob)(\Phib(\xb)-\Phib(\zerob)+\Phib(\zerob))\| 
        \nonumber\\
        &\quad + 
        \Big\| \sum_{i=1}^m u_i ({P_\bbV\cK_{\Delta t}^{\eb_i}|_\bbV} - \cK_{\Delta t,d}^{\eb_i})(\Phib(\xb)-\Phib(\zerob)+\Phib(\zerob)) \Big\|
        \nonumber\\
        &\leq \Big( 
            \hat{u} \|{P_\bbV\cK_{\Delta t}^\zerob|_\bbV} - \cK_{\Delta t,d}^\zerob\| 
        \nonumber\\
        &\qquad\quad\quad
        + \sum_{i=1}^m |u_i|\|{P_\bbV\cK_{\Delta t}^{\eb_i}|_\bbV} - \cK_{\Delta t,d}^{\eb_i}\| 
        \Big)
        \|\Phib(\xb)-\Phib(\zerob)\|
        \nonumber\\
        &\; + 
        \hat{u} \|({P_\bbV\cK_{\Delta t}^\zerob|_\bbV} - \cK_{\Delta t,d}^\zerob)\Phib(\zerob)\| \\ 
        &\; +
        \left\|
            \left[\begin{smallmatrix}
                ({P_\bbV\cK_{\Delta t}^{\eb_1}|_\bbV})_{21} - (\cK_{\Delta t,d}^{\eb_1})_{21} & \cdots & ({P_\bbV\cK_{\Delta t}^{\eb_m}|_\bbV})_{21} - (\cK_{\Delta t,d}^{\eb_m})_{21}
            \end{smallmatrix}\right]
        \right\| \|\ub\|.
    \end{align*}
    \normalsize
    Next, we exploit $({P_\bbV\cK_{\Delta t}^\zerob|_\bbV} - \cK_{\Delta t,d}^\zerob)\Phib(\zerob) = \zerob$ due to the structure of the Koopman operator in~\eqref{eq:structure-Koopman-operator} and its data-driven approximate in~\eqref{eq:structure-Koopman-operator-data}, $\|{P_\bbV\cK^{\bar{\ub}}_{\Delta t}|_\bbV} - \cK^{\bar{\ub}}_{\Delta t,d}\|\leq c_\eta$ for $\bar{\ub}\in\{\zerob,\eb_1,...,\eb_m\}$ according to~\eqref{eq:error-bound-operator}, and%
    \vspace*{-2.25\baselineskip}
    \small%
    \begin{align*}
        &\left\|
            \begin{bmatrix}
                ({P_\bbV\cK_{\Delta t}^{\eb_1}|_\bbV})_{21} - (\cK_{\Delta t,d}^{\eb_1})_{21} 
                & \cdots & 
                ({P_\bbV\cK_{\Delta t}^{\eb_m}|_\bbV})_{21} - (\cK_{\Delta t,d}^{\eb_m})_{21}
            \end{bmatrix}
        \right\|
        \\
        &\leq \sqrt{
            \sum_{i=1}^m
            \left\|
                ({P_\bbV\cK_{\Delta t}^{\eb_i}|_\bbV})_{21} - (\cK_{\Delta t,d}^{\eb_i})_{21}
            \right\|^2
        } \\
        &\leq \sqrt{
            \sum_{i=1}^m
            \left\|
                {P_\bbV\cK_{\Delta t}^{\eb_i}|_\bbV} - \cK_{\Delta t,d}^{\eb_i}
            \right\|^2
        },
    \end{align*}
    \normalsize
    which results in~\eqref{eq:proportional-bound-DD-surrogate} with
    \begin{equation}\label{eq:error-estimate-c3-c4}
        c_3 = \hat{u} + \max_{\ub\in\bbU} \| \ub \|_1.
    \end{equation}
    This completes the proof.
\end{pf}
Again, we emphasize that the constant $c_3$ in Lemma~\ref{lm:proportional-bound-DD-surrogate} can be evaluated for a given set $\bbU$.

Now we can prove Theorem~\ref{thm:proportional-bound-data}.

\textbf{PROOF} (Theorem~\ref{thm:proportional-bound-data})\textbf{.}\;
    This is a direct consequence of Lemma~\ref{lm:proportional-bound-bilinear-representation} and Lemma~\ref{lm:proportional-bound-DD-surrogate}. 
    We observe $({P_\bbV\cK^\ub_{\Delta t}|_\bbV}-\cK^\ub_{\Delta t,d})\Phib(\xb)=\xib(\xb,\ub)+\etab(\xb,\ub)$ and use the estimates in~\eqref{eq:proportional-bound-bilinear-representation} and~\eqref{eq:proportional-bound-DD-surrogate} to deduce
    \begin{align*}
        &\|({P_\bbV\cK^\ub_{\Delta t}|_\bbV}-\cK^\ub_{\Delta t,d})\Phib(\xb)\| 
        \leq \|\xib(\xb,\ub)\| + \|\etab(\xb,\ub)\|
        \\
        &\leq (\Delta t^2 c_1 + c_3c_\eta)\|\Phib(\xb)-\Phib(\zerob)\| + \sqrt{m}(\Delta t^2 c_2 + c_\eta) \|\ub\|.
    \\[-1.5\baselineskip]
    \end{align*}
    Then, defining
    \begin{equation}\label{eq:LxLu}
    \hspace*{-0.03\linewidth}
        \bar{c}_x=\Delta t^2c_1 + c_3 c_\eta \;\;\text{and}\;\; \bar{c}_u=\sqrt{m}(\Delta t^2 c_2+c_\eta).
    \end{equation}
    results in~\eqref{eq:proportional-bound-data}. 
    In view of Lemma~\ref{lm:proportional-bound-DD-surrogate}, and for any probabilistic tolerance $\delta \in (0,1)$, the constant $c_\eta$ is of order $c_\eta = \mathcal{O}\left(\nicefrac{1}{\sqrt{\delta d_0}}\right)$ which implies the claim $\bar{c}_x,\bar{c}_u = \mathcal{O}\left(\nicefrac{1}{\sqrt{\delta d_0}} + \Delta t^2\right)$.
\null\hfill$\square$

%% file: secAppendix-proof-controller-design.tex
\section{Proof of Theorem~\ref{thm:stability-condition-LFR}}\label{app:proof-thm-stability-condition-LFR}

In the following, we prove Theorem~\ref{thm:stability-condition-LFR} guaranteeing closed-loop exponential stability of the sampled nonlinear system~\eqref{eq:dynamics-nonlinear-sampled} with the feedback control law~\eqref{eq:controller-parametrization-nonlinear}.
To this end, we first observe that the inclusion of the constant observable $\phi_0\equiv 1$ and the thereof resulting structure of the Koopman operator~\eqref{eq:structure-Koopman-operator} and its data-driven approximates~\eqref{eq:structure-Koopman-operator-data} yields the data-driven surrogate model
\begin{align*}
   \Phib(\xb_{k+1}) 
   &= \cK^\ub_{\Delta t,d} \Phib(\xb) + (\cK_{\Delta t}^\ub - \cK_{\Delta t,d}^\ub)\Phib(\xb)
   \\
   &= \cK_{\Delta t,d}^\zerob \Phib(\xb_k) 
   \\
   &\;\;
   + \sum_{i=1}^m (\ub_k)_i(\cK_{\Delta t,d}^{\eb_i} - \cK_{\Delta t,d}^\zerob)\Phib(\xb_k) 
   + \zetab(\xb_k, \ub_k)
\end{align*}
with $\zetab(\xb,\ub)\coloneqq(\cK_{\Delta t,d}^\ub - \cK_{\Delta t}^\ub)\Phib(\xb)$.
In particular, we observe that the first element of the learning error $\zetab$ is zero and, thus, $\|\zetab(\xb,\ub)\| = \|\hat{\zetab}(\xb,\ub)\|$ with $\hat{\zetab}(\xb,\ub)\coloneqq \begin{bmatrix} \zerob_{N\times 1} & I_N \end{bmatrix}\zetab(\xb,\ub)$. 
Further, we exploit the structure in~\eqref{eq:structure-Koopman-operator},~\eqref{eq:structure-Koopman-operator-data} to deduce the lifted representation
\begin{equation}\label{eq:dynamics-lifted-operator}
    \hat{\Phib}(\xb_{k+1})
    = A \hat{\Phib}(\xb_k) 
    + B_0 \ub_k
    + \tB(\ub_k\kron \hat{\Phib}(\xb_k)) 
    + \hat{\zetab}(\xb,\ub).
\end{equation}
Here, we use the reduced observable function $\hat{\Phib}(\xb)=\begin{bmatrix}\zerob_{N\times 1} & I_N\end{bmatrix}\Phib(\xb)$, where we removed the constant observable $\phi_0\equiv 1$ such that $\hat{\Phib}(\zerob)=\zerob$.
According to the presented bilinear surrogate in Algorithm~\ref{alg:learning-architecture} satisfying the error bounds in Corollary~\ref{cor:proportional-bound-error}, the reduced approximation error $\hat{\zetab}(\xb,\ub)$ satisfies
\begin{equation}\label{eq:proportional-bound-remainder}
    \|\hat{\zetab}(\xb,\ub)\| 
    = \|\zetab(\xb,\ub)\|
    \leq 
    c_x \|\hat{\Phib}(\xb)\| + c_u \|\ub\|
\end{equation}
for all $\xb\in\bbX$ and $\ub\in\bbU$.

Inspired by the controller designs in~\cite{strasser:berberich:allgower:2023a,strasser:berberich:allgower:2023b,strasser:schaller:worthmann:berberich:allgower:2025}, we view~\eqref{eq:dynamics-lifted-operator} as an uncertain bilinear system with uncertainty $\hat{\zetab}(\xb,\ub)$. 
In particular, we use linear robust control techniques to cope with the uncertainty via convex optimization. 
More precisely, we define the remainder $\varepsilonb:\bbR^N\times\bbR^m\to\bbR^N$ depending on the lifted state as
\begin{equation}\label{eq:epsilon-definition}
    \varepsilonb(\vb_1,\vb_2)=\hat{\zetab}(\begin{bmatrix}I_n&0_{n\times N-n}\end{bmatrix}\vb_1,\vb_2).
\end{equation}
According to the error bound~\eqref{eq:proportional-bound-remainder}, $\varepsilonb$ satisfies
\begin{equation}\label{eq:proportional-bound-epsilon}
    \|\varepsilonb(\hat{\Phib}(\xb),\ub)\| = \|\zetab(\xb,\ub)\| \leq c_x\|\hat{\Phib}(\xb)\| + c_u\|\ub\|
\end{equation}
for all $\xb\in\bbX$ and $\ub\in\bbU$.
To handle the nonlinear term $(\ub\kron\hat{\Phib}(\xb))$, we introduce an artificial uncertainty $\psib\in\mathbf{\Delta}_{\Phib}$ over-approximating $\hat{\Phib}(\xb)$.
This uncertainty needs to be bounded, i.e., we ensure $\hat{\Phib}(\xb)\in\mathbf{\Delta}_{\Phib}$ for all times.
Then, we stabilize
\begin{equation*}
    \hat{\Phib}(\xb_{k+1}) = A \hat{\Phib}(\xb_k) + B_0 \ub_k + \tB(I_m\kron \psib)\ub_k + \varepsilonb(\hat{\Phib}(\xb_k),\ub_k)
\end{equation*}
for all $\psib\in\mathbf{\Delta}_{\Phib}$ and perturbation functions $\varepsilonb$ satisfying~\eqref{eq:proportional-bound-epsilon} for all $\xb\in\bbX,\ub\in\bbU$.
Setting $A_K = A+B_0K$, $B_{K_w}=\tB + B_0K_w$, and substituting the input by the feedback $\ub=\mub_{\psib}(\xb)=K\hat{\Phib}(\xb)+K_w(I_m\kron\psib)\mub_{\psib}(\xb)$, we obtain the corresponding closed-loop system 
\begin{multline}\label{eq:dynamics-lifted-operator-Delta-closed-loop}
    \hat{\Phib}(\xb_{k+1}) 
    = A_K \hat{\Phib}(\xb_k) 
    + B_{K_w}(I_m\kron\psib) \mub_{\psib}(\xb)
    \\
    + \varepsilonb(\hat{\Phib}(\xb),\mub_{\psib}(\xb))
\end{multline}
with $\psib\in\mathbf{\Delta}_{\Phib}$ and $\varepsilonb$ satisfying~\eqref{eq:proportional-bound-epsilon} for all $\xb\in\bbX$.

The remainder of the proof of Theorem~\ref{thm:stability-condition-LFR} proceeds similarly to~\cite[Thm.~7]{strasser:schaller:worthmann:berberich:allgower:2025} with the key difference of discrete vs. continuous time, where the main adaptations are highlighted in the following. 
\begin{pf}    
    The presented proof is separated into two parts. 
    First, we show that all $\xb\in\cX_\mathrm{RoA}$ satisfy $\hat{\Phib}(\xb)\in\mathbf{\Delta}_{\Phib}$.
    Second, we conclude positive invariance of $\cX_\mathrm{RoA}$ together with exponential stability of the sampled closed-loop system~\eqref{eq:dynamics-nonlinear-sampled} for all $\hat{\xb}\in\cX_\mathrm{RoA}$.
    
    \textit{Part I: $\xb\in\cX_\mathrm{RoA}$ implies $\hat{\Phib}(\xb)\in\mathbf{\Delta}_{\Phib}$:~}
    In order to represent the lifted dynamics via~\eqref{eq:dynamics-lifted-operator-Delta-closed-loop}, we require $\hat{\Phib}(\xb)\in\mathbf{\Delta}_{\Phib}$ for all times.
    To this end, we exploit that
    \begin{align*}
        0 
        &\succeq
        \frac{1}{\nu}
        \begin{bmatrix}
            \nu\tr_z - 1 & -\nu\tsb_z^\top \\ -\nu\tsb_z & \nu\tQ_z + P 
        \end{bmatrix}
        \\
        &= 
        \left[\star\right]^\top
        \begin{bmatrix}
            \tQ_z & \tsb_z \\ \tsb_z^\top & \tr_z
        \end{bmatrix}
        \begin{bmatrix}
            0 & I \\ -1 & 0
        \end{bmatrix}
        + 
        \left[\star\right]^\top
        \begin{bmatrix}
            -\frac{1}{\nu} & 0 \\ 0 & \frac{1}{\nu}P
        \end{bmatrix}
        \begin{bmatrix}
            1 & 0 \\ 0 & -I
        \end{bmatrix}
    \end{align*}     
    is equivalent to~\eqref{eq:stability-condition-LFR-invariance} after dividing the inequality by $\nu$.
    We rewrite this inequality as
    \begin{equation*}
        \left[\def\arraystretch{1.15}\begin{array}{cc}
            0 & I \\ -1 & 0 \\\hline
            1 & 0 \\ 0 & -I
        \end{array}\right]^\top 
        \left[\def\arraystretch{1.15}\begin{array}{cc|cc}
            \tQ_z & \tsb_z & 0 & 0 \\
            \tsb_z^\top & \tr_z & 0 & 0 \\\hline
            0 & 0 & -\frac{1}{\nu} & 0 \\
            0 & 0 & 0 & \frac{1}{\nu}P
        \end{array}\right]
        \left[\def\arraystretch{1.15}\begin{array}{cc}
            0 & I \\ -1 & 0 \\\hline
            1 & 0 \\ 0 & -I
        \end{array}\right]
        \preceq 0
    \end{equation*}
    and apply the dualization lemma~\citep[Lm. 4.9]{scherer:weiland:2000}.
    As a consequence, we obtain
    \begin{equation*}
        \left[\def\arraystretch{1.15}\begin{array}{cc}
            I & 0 \\ 0 & 1 \\\hline
            0 & 1 \\ I & 0
        \end{array}\right]^\top 
        \left[\def\arraystretch{1.15}\begin{array}{cc|cc}
            Q_z & \sb_z & 0 & 0 \\
            \sb_z^\top & r_z & 0 & 0 \\\hline
            0 & 0 & -\nu & 0 \\
            0 & 0 & 0 & \nu P^{-1}
        \end{array}\right]
        \left[\def\arraystretch{1.15}\begin{array}{cc}
            I & 0 \\ 0 & 1 \\\hline
            0 & 1 \\ I & 0
        \end{array}\right]
        \preceq 0,
    \end{equation*}
    i.e., 
    \begin{equation*}
        \begin{bmatrix}
            Q_z & \sb_z \\ \sb_z^\top & r_z
        \end{bmatrix}
        - \nu \begin{bmatrix}
            - P^{-1} & 0 \\ 0 & 1
        \end{bmatrix}
        \succeq 0.
    \end{equation*}
    Hence, we recall the definition of $\mathbf{\Delta}_{\Phib}$ in~\eqref{eq:DeltaPhi-definition} and deduce $\hat{\Phib}(\xb)\in\mathbf{\Delta}_{\Phib}$ for all $\xb\in\cX_\mathrm{RoA}$ by the multiplication from left and right by $
        \begin{bmatrix}
            \hat{\Phib}(\xb)^\top & 1
        \end{bmatrix}^\top
    $ and its transpose, respectively, for $\xb\in\cX_\mathrm{RoA}$ (cf.\ the S-procedure~\citep{scherer:weiland:2000,boyd:vandenberghe:2004}).

    \textit{Part II: Positive invariance of $\cX_\mathrm{RoA}$ and closed-loop exponential stability:~}
    In the following, we show $\xb_+\in\cX_\mathrm{RoA}$ for all $\xb\in\cX_\mathrm{RoA}$ to conclude positive invariance of $\cX_\mathrm{RoA}$, where we use $(\xb_+,\xb)$ as a short-hand notation for $(\xb_{k+1},\xb_k)$. 
    The obtained RoA $\cX_\mathrm{RoA}$ will then result as a Lyapunov sublevel set, for which we define the Lyapunov function candidate $V(\xb) = \hat{\Phib}(\xb)^\top P^{-1} \hat{\Phib}(\xb)$, i.e., positive invariance can be inferred if $\Delta V(\xb)=V(\xb_+)-V(\xb) \leq 0$ for all $\xb\in\cX_\mathrm{RoA}$.

    Recall $A_K = A + B_0 K$ and $B_{K_w}=\tB+B_0K_w$.
    Then, the definitions $K = L P^{-1}$ and $K_w=L_w(\Lambda^{-1}\kron I_N)$ together with twice applying the Schur complement~\citep[cf.][]{boyd:vandenberghe:2004} to~\eqref{eq:stability-condition-LFR} and using the dualization lemma~\citep[Lm. 4.9]{scherer:weiland:2000} (compare the arguments in the proof of~\cite[Thm.~7]{strasser:schaller:worthmann:berberich:allgower:2025}) yields%
    \scriptsize
    \begin{equation}\label{eq:proof-primal}
        \tilde{\Psi}^\top \diag\left(
            \begin{bmatrix}
                -P^{-1} & 0 \\ 0 & P^{-1}
            \end{bmatrix},
            \begin{bmatrix}
                \tilde{\Lambda}\kron{Q_z} & \tilde{\Lambda}\kron{\sb_z} \\ \tilde{\Lambda}\kron{\sb_z^\top} & \tilde{\Lambda}\kron{r_z}
            \end{bmatrix},
            \tau^{-1} \Pi_r
        \right)
        \tilde{\Psi} 
        \prec 0,
    \end{equation}
    \normalsize
    with $\tilde{\Lambda}=\Lambda^{-1}$,
    $
        \Pi_r = 
        \left[\begin{smallmatrix}
            -I_N & 0 \\ 0 & 2\left[\begin{smallmatrix} c_x^2 I_N & 0 \\ 0 & c_u^2I_m \end{smallmatrix}\right]
        \end{smallmatrix}\right]
    $, and
    \begin{equation}
        \tilde{\Psi}^\top 
        = \left[\begin{array}{cc|cc|cc}
            I & A_K^\top & 0 & K^\top & 0 & \begin{bmatrix}I&K^\top\end{bmatrix} \\
            0 & B_{K_w}^\top & I & K_w^\top & 0 & \begin{bmatrix}0&K_w^\top\end{bmatrix} \\
            0 & I & 0 & 0 & I & 0
        \end{array}\right].
    \end{equation}
    Then, we multiply~\eqref{eq:proof-primal} from the left and from the right by $
        \begin{bmatrix}
            \hat{\Phib}(\xb)^\top & (\mub_{\psib}(\xb)\kron\psib)^\top & \varepsilonb(\hat{\Phib}(\xb),\mub_{\psib}(\xb))^\top
        \end{bmatrix}^\top
    $ and its transpose, respectively, where $\xb\in\cX_\mathrm{RoA}$, $\mub_{\psib}(\xb)=K\hat{\Phib}(\xb)+K_w(\mub_{\psib}(\xb)\kron\psib)$, $\psib\in\mathbf{\Delta}_{\Phib}$, and $\varepsilonb(\hat{\Phib}(\xb),\mub_{\psib}(\xb))$ satisfies~\eqref{eq:proportional-bound-epsilon}.
    As similarly shown in the proof of~\cite[Thm.~7]{strasser:schaller:worthmann:berberich:allgower:2025} for continuous time, exploiting the S-procedure~\citep{scherer:weiland:2000,boyd:vandenberghe:2004} leads to
    \vspace*{-2\baselineskip}\small
    \begin{multline}\label{eq:proof-nonlinear-dissipation-inequality-substitute}
        (\star)^\top P^{-1} \Big(A_K \hat{\Phib}(\xb) + B_{K_w}(\mub_{\psib}(\xb)\kron\psib) + \varepsilonb(\hat{\Phib}(\xb),\mub_{\psib}(\xb))\Big)
        \\
        - \hat{\Phib}(\xb)^\top P^{-1} \hat{\Phib}(\xb) 
        < 0
    \end{multline}
    \normalsize
    if $\xb\in\cX_\mathrm{RoA}\setminus\{\zerob\}$, $\psib\in\mathbf{\Delta}_{\Phib}$, and if $\varepsilonb(\hat{\Phib}(\xb),\mub_{\psib}(\xb))$ satisfies~\eqref{eq:proportional-bound-epsilon}.
    
    Now, with the Lyapunov candidate function $V(x)$, we substitute~\eqref{eq:dynamics-lifted-operator-Delta-closed-loop} in~\eqref{eq:proof-nonlinear-dissipation-inequality-substitute} to deduce
    \begin{equation}\label{eq:proof-Lyapunov-decay}
        \Delta V(\xb)
        = \hat{\Phib}(\xb_+)^\top P^{-1} \hat{\Phib}(\xb_+)       
        - \hat{\Phib}(\xb)^\top P^{-1} \hat{\Phib}(\xb) 
        < 0
    \end{equation}
    if $\xb\in\cX_\mathrm{RoA}\setminus\{\zerob\}$, $\psib\in\mathbf{\Delta}_{\Phib}$, and if $\varepsilonb(\hat{\Phib}(\xb),\mub_{\psib}(\xb))$ satisfies~\eqref{eq:proportional-bound-epsilon}.
    Hence, the controller $\mub_{\psib}(\xb)=K\hat{\Phib}(\xb)+K_w(\mub_{\psib}(\xb)\kron\Delta_{\Phib})$ guarantees~\eqref{eq:proof-Lyapunov-decay} for solutions of the lifted system~\eqref{eq:dynamics-lifted-operator-Delta-closed-loop} if $\xb\in\cX_\mathrm{RoA}\setminus\{\zerob\}$, $\psib\in\mathbf{\Delta}_{\Phib}$, and if $\varepsilonb(\hat{\Phib}(\xb),\mub_{\psib}(\xb))$ satisfies~\eqref{eq:proportional-bound-epsilon}.
    As proven in Part I, we have $\hat{\Phib}(\xb)\in\mathbf{\Delta}_{\Phib}$ for all $\xb\in\cX_\mathrm{RoA}$.
    Further, $\varepsilonb(\hat{\Phib}(\xb),\mub(\xb))=\hat{\zetab}(\xb,\mub(\xb))$ due to~\eqref{eq:epsilon-definition}.
    Consequently, the controller 
    \begin{align*}
        \mub(\xb)
        &=\mub_{\psib=\hat{\Phib}(\xb)}(\xb)
        =K\hat{\Phib}(\xb)+K_w(I_m\kron\hat{\Phib}(\xb))\mub(\xb)
        \\
        &=(I-K_w(I_m\kron\hat{\Phib}(\xb)))^{-1}K\hat{\Phib}(\xb)
    \end{align*}
    renders $\cX_\mathrm{RoA}$ positive invariant by guaranteeing~\eqref{eq:proof-Lyapunov-decay} in particular for system~\eqref{eq:dynamics-lifted-operator} for all $\xb\in\cX_\mathrm{RoA}\setminus\{\zerob\}$ if $\hat{\zetab}(\xb,\mub(\xb))$ satisfies~\eqref{eq:proportional-bound-remainder}.
    Moreover, we deduce robust exponential stability of~\eqref{eq:dynamics-lifted-operator} for all $\xb\in \cX_\mathrm{RoA}$ if $\hat{\zetab}(\xb,\mub(\xb))$ satisfies~\eqref{eq:proportional-bound-remainder}.
    
    Finally, we exploit the error bounds of SafEDMD in Corollary~\ref{cor:proportional-bound-error}, i.e., we deduce that the data-driven surrogates $K^{\bar{\ub}}_{\Delta t,d}$, $\bar{\ub}\in\{\zerob,\eb_1,...,\eb_m\}$, and the resulting error $\hat{\zetab}(\xb,\mub(\xb))$ of the learning scheme in Section~\ref{sec:learning-architecture} satisfy the obtained error bound~\eqref{eq:proportional-bound-error} for a data length $d\geq d_0$ with probability $1-\delta$. 
    Hence, we conclude closed-loop exponential stability of the sampled nonlinear system~\eqref{eq:dynamics-nonlinear-sampled} for all $\xb\in\cX_\mathrm{RoA}$ with probability $1-\delta$.
\end{pf}

%% file: secAppendix-proof-controller-continuous-time.tex
\section{Proof of Corollary~\ref{cor:controller-certificates-continuous-time}}\label{app:proof-controller-continuous-time}
In the following, we prove the stability guarantees in Corollary~\ref{cor:controller-certificates-continuous-time} for the continuous-time system~\eqref{eq:dynamics-nonlinear} in closed loop with the sampling-based controller~\eqref{eq:controller-sampled}.
\begin{pf}  
    In order to show the statement, we combine Theorem~\ref{thm:stability-condition-LFR} with arguments used in~\cite[Thm.~2.27]{grune:pannek:2017} to infer stability guarantees for the continuous-time system~\eqref{eq:dynamics-nonlinear} based on the sampled controller~\eqref{eq:controller-sampled}.
    In particular, we need the solutions of the sampled-data closed-loop system to be uniformly bounded over $\Delta t$ as defined in~\cite[Def.~2.24]{grune:pannek:2017}, i.e., we have to show the existence of a $\cK$-function\footnote{We denote the class of functions $\alpha\!:\bbR_{\geq 0} \to \bbR_{\geq 0}$, which are continuous, strictly increasing, and satisfy $\alpha(0)=0$ by $\cK$.} $\gamma$ such that for all $\hat{\xb}\in\cX_\mathrm{RoA}$, the solutions of the nonlinear system~\eqref{eq:dynamics-nonlinear} satisfy
    \begin{equation}\label{eq:proof-continuous-time-uniform-bound}
        \|\xb(\tau;\hat{\xb},\mub_\mathrm{s}(\xb))\| \leq \gamma(\|\hat{\xb}\|)
    \end{equation}
    for all $\tau\in[0,\Delta t]$.
    Note that the sampled controller $\mub_s(x)$ yields a constant value $\mub_s(\xb(\tau))\equiv\bar{\mub}\in\bbU$ for $\tau\in[0,\Delta t)$.
    Thus, we observe
    \begin{multline*}
        \| \xb(\tau;\hat{\xb},\bar{\mub}) \| 
        \leq \max_{\bar{\ub} \in \bbU} \| \xb(\tau;\hat{\xb},\ub) \| 
        \\
        \overset{\eqref{eq:philowerupper}} {\leq} \max_{\bar{\ub} \in \bbU} \| \Phib(\xb(\tau;\hat{\xb},\ub)) - \Phib(\zerob) \| 
    \end{multline*}
    for all $\tau \in [0,\Delta t)$ with $\ub(t) \equiv \bar{\ub}$ on $[0,\Delta t)$.
    Further, recalling $\phi_0\equiv 1$ and, hence, $(\cK_\tau^{\ub}\Phib)(\zerob)=\Phib(\zerob)$, we deduce
    \begin{align*}
        \Phib(\xb(\tau;\hat{\xb},\ub)) - \Phib(\zerob)
        \overset{\eqref{eq:Koopman-operator}}&{=} (\cK_{\tau}^{\ub} \Phib) (\hat{\xb}) - (\cK_{\tau}^{\ub} \Phib)(\zerob) 
        \\
        &= e^{\tau\cL^{\bar{\ub}}} (\Phib( \hat{\xb} ) - \Phib(\zerob))
    \end{align*}
    for all $\tau \in [0,\Delta t)$.
    Then,
    \begin{equation*}
         \| \xb(\tau;\hat{\xb},\bar{\mub}) \| 
        \leq e^{ \Delta t \| \mathcal{L}^\star \| } \|\Phib(\hat{\xb}) - \Phib(\zerob) \|
        \overset{\eqref{eq:philowerupper}}{\leq} e^{ \Delta t \| \mathcal{L}^\star \| } L_\Phi \| \hat{\xb} \|
    \end{equation*}
    where $\| \mathcal{L}^\star \| \coloneqq \max_{\bar{\ub} \in \mathbb{U}} \| \mathcal{L}^{\bar{\ub}} \|$. 
    Hence,~\eqref{eq:proof-continuous-time-uniform-bound} holds for the linear function $\gamma(r)\coloneqq c r$ with $c \coloneqq e^{ \Delta t \| \mathcal{L}^\star \| } L_\Phi$.
    Moreover, exponential stability of the sampled system guarantees the existence of $C \geq 1$ and $\sigma\in(0,1)$ such that $\|\xb(k\Delta t)\|\leq C \sigma^k \|\hat{\xb}\|$ holds for all $k\geq 0$, which corresponds to the $\cK\cL$-function\footnote{A continuous function $\beta: \bbR_{\geq 0}\times\bbR_{\geq 0}\to\bbR_{\geq 0}$ belongs to class $\cK\cL$ if $\beta(\cdot,s)\in\cK$ for each fixed $s$ and $\beta(r,\cdot)$ is decreasing with $\lim_{s_\to\infty} \beta(r,s)=0$ for each fixed $r$.} $\beta(r,k) = C \sigma^kr$. 
    Hence, for any $t\in[k\Delta t, (k+1)\Delta t)$ with $k\geq 0$ we get
    \begin{align*}
        \|\xb(t;\hat{\xb},\mub_s(\xb))\| 
        &\leq \gamma(\|\xb(k\Delta t;\hat{\xb},\bar{\mub}_k)\|) 
        \leq \gamma(\beta(\|\hat{\xb}\|,k)) 
        \\
        &\leq c C \sigma^k \| \hat{\xb} \| 
        \leq \widetilde{C} e^{ - t \eta} \| \hat{\xb} \|
    \end{align*}
    for $\bar{\mub}_k=\mub_\mathrm{s}(\xb(k\Delta t))$ and with $\eta := - \ln(\sigma) \Delta t^{-1}>0$ and $\widetilde{C} := c C \sigma^{-1}$, i.e., exponential stability of the continuous-time closed-loop system.
\end{pf}